\documentclass[longauth]{aa}

\usepackage{graphicx}
\usepackage{natbib}
\usepackage{scalerel}

\usepackage[table, dvipsnames]{xcolor}

\usepackage{txfonts}
\usepackage[pdfencoding=auto,psdextra]{hyperref}
\hypersetup{
    colorlinks=true,
    linkcolor=blue,
    filecolor=magenta,
    urlcolor=blue,
    citecolor=blue
}
\makeatletter
\renewcommand*\aa@pageof{, page \thepage{} of \pageref*{LastPage}}
\makeatother

\usepackage[utf8]{inputenc}

\usepackage{etoolbox}
\makeatletter
\apptocmd{\appendix}{\nolinenumbers}{}{}
\makeatother

\usepackage{euclid}
\usepackage{macros}
\usepackage{multirow}
\usepackage{pifont}

\begin{document}

\title{\Euclid preparation}
\subtitle{Three-dimensional galaxy clustering in configuration space: 
Three-point correlation function estimation}    

\newcommand{\orcid}[1]{} 
\author{Euclid Collaboration: A.~Veropalumbo\orcid{0000-0003-2387-1194}\thanks{\email{alfonso.veropalumbo@inaf.it}}\inst{\ref{aff1},\ref{aff2},\ref{aff3}}
\and M.~Moresco\orcid{0000-0002-7616-7136}\inst{\ref{aff4},\ref{aff5}}
\and F.~Marulli\orcid{0000-0002-8850-0303}\inst{\ref{aff4},\ref{aff5},\ref{aff6}}
\and E.~Branchini\orcid{0000-0002-0808-6908}\inst{\ref{aff3},\ref{aff2},\ref{aff1}}
\and M.~Guidi\orcid{0000-0001-9408-1101}\inst{\ref{aff7},\ref{aff5}}
\and A.~Farina\orcid{0009-0000-3420-929X}\inst{\ref{aff3},\ref{aff1},\ref{aff2}}
\and A.~Pugno\inst{\ref{aff8}}
\and E.~Sefusatti\orcid{0000-0003-0473-1567}\inst{\ref{aff9},\ref{aff10},\ref{aff11}}
\and D.~Tavagnacco\orcid{0000-0001-7475-9894}\inst{\ref{aff9}}
\and F.~Rizzo\orcid{0000-0002-9407-585X}\inst{\ref{aff9}}
\and E.~Romelli\orcid{0000-0003-3069-9222}\inst{\ref{aff9}}
\and S.~de~la~Torre\inst{\ref{aff12}}
\and A.~Eggemeier\orcid{0000-0002-1841-8910}\inst{\ref{aff8}}
\and E.~Sihvola\orcid{0000-0003-1804-7715}\inst{\ref{aff13}}
\and M.~Viel\orcid{0000-0002-2642-5707}\inst{\ref{aff10},\ref{aff9},\ref{aff14},\ref{aff11},\ref{aff15}}
\and N.~Aghanim\orcid{0000-0002-6688-8992}\inst{\ref{aff16}}
\and B.~Altieri\orcid{0000-0003-3936-0284}\inst{\ref{aff17}}
\and S.~Andreon\orcid{0000-0002-2041-8784}\inst{\ref{aff1}}
\and N.~Auricchio\orcid{0000-0003-4444-8651}\inst{\ref{aff5}}
\and C.~Baccigalupi\orcid{0000-0002-8211-1630}\inst{\ref{aff10},\ref{aff9},\ref{aff11},\ref{aff14}}
\and M.~Baldi\orcid{0000-0003-4145-1943}\inst{\ref{aff7},\ref{aff5},\ref{aff6}}
\and S.~Bardelli\orcid{0000-0002-8900-0298}\inst{\ref{aff5}}
\and P.~Battaglia\orcid{0000-0002-7337-5909}\inst{\ref{aff5}}
\and A.~Biviano\orcid{0000-0002-0857-0732}\inst{\ref{aff9},\ref{aff10}}
\and M.~Brescia\orcid{0000-0001-9506-5680}\inst{\ref{aff18},\ref{aff19}}
\and S.~Camera\orcid{0000-0003-3399-3574}\inst{\ref{aff20},\ref{aff21},\ref{aff22}}
\and G.~Ca\~nas-Herrera\orcid{0000-0003-2796-2149}\inst{\ref{aff23},\ref{aff24}}
\and V.~Capobianco\orcid{0000-0002-3309-7692}\inst{\ref{aff22}}
\and C.~Carbone\orcid{0000-0003-0125-3563}\inst{\ref{aff25}}
\and V.~F.~Cardone\inst{\ref{aff26},\ref{aff27}}
\and J.~Carretero\orcid{0000-0002-3130-0204}\inst{\ref{aff28},\ref{aff29}}
\and S.~Casas\orcid{0000-0002-4751-5138}\inst{\ref{aff30},\ref{aff31}}
\and M.~Castellano\orcid{0000-0001-9875-8263}\inst{\ref{aff26}}
\and G.~Castignani\orcid{0000-0001-6831-0687}\inst{\ref{aff5}}
\and S.~Cavuoti\orcid{0000-0002-3787-4196}\inst{\ref{aff19},\ref{aff32}}
\and K.~C.~Chambers\orcid{0000-0001-6965-7789}\inst{\ref{aff33}}
\and A.~Cimatti\inst{\ref{aff34}}
\and C.~Colodro-Conde\inst{\ref{aff35}}
\and G.~Congedo\orcid{0000-0003-2508-0046}\inst{\ref{aff23}}
\and C.~J.~Conselice\orcid{0000-0003-1949-7638}\inst{\ref{aff36}}
\and L.~Conversi\orcid{0000-0002-6710-8476}\inst{\ref{aff37},\ref{aff17}}
\and Y.~Copin\orcid{0000-0002-5317-7518}\inst{\ref{aff38}}
\and F.~Courbin\orcid{0000-0003-0758-6510}\inst{\ref{aff39},\ref{aff40},\ref{aff41}}
\and H.~M.~Courtois\orcid{0000-0003-0509-1776}\inst{\ref{aff42}}
\and A.~Da~Silva\orcid{0000-0002-6385-1609}\inst{\ref{aff43},\ref{aff44}}
\and H.~Degaudenzi\orcid{0000-0002-5887-6799}\inst{\ref{aff45}}
\and G.~De~Lucia\orcid{0000-0002-6220-9104}\inst{\ref{aff9}}
\and H.~Dole\orcid{0000-0002-9767-3839}\inst{\ref{aff16}}
\and F.~Dubath\orcid{0000-0002-6533-2810}\inst{\ref{aff45}}
\and X.~Dupac\inst{\ref{aff17}}
\and S.~Dusini\orcid{0000-0002-1128-0664}\inst{\ref{aff46}}
\and S.~Escoffier\orcid{0000-0002-2847-7498}\inst{\ref{aff47}}
\and M.~Farina\orcid{0000-0002-3089-7846}\inst{\ref{aff48}}
\and R.~Farinelli\inst{\ref{aff5}}
\and F.~Faustini\orcid{0000-0001-6274-5145}\inst{\ref{aff26},\ref{aff49}}
\and S.~Ferriol\inst{\ref{aff38}}
\and F.~Finelli\orcid{0000-0002-6694-3269}\inst{\ref{aff5},\ref{aff50}}
\and P.~Fosalba\orcid{0000-0002-1510-5214}\inst{\ref{aff51},\ref{aff52}}
\and S.~Fotopoulou\orcid{0000-0002-9686-254X}\inst{\ref{aff53}}
\and M.~Frailis\orcid{0000-0002-7400-2135}\inst{\ref{aff9}}
\and E.~Franceschi\orcid{0000-0002-0585-6591}\inst{\ref{aff5}}
\and M.~Fumana\orcid{0000-0001-6787-5950}\inst{\ref{aff25}}
\and S.~Galeotta\orcid{0000-0002-3748-5115}\inst{\ref{aff9}}
\and K.~George\orcid{0000-0002-1734-8455}\inst{\ref{aff54}}
\and W.~Gillard\orcid{0000-0003-4744-9748}\inst{\ref{aff47}}
\and B.~Gillis\orcid{0000-0002-4478-1270}\inst{\ref{aff23}}
\and C.~Giocoli\orcid{0000-0002-9590-7961}\inst{\ref{aff5},\ref{aff6}}
\and P.~G\'omez-Alvarez\orcid{0000-0002-8594-5358}\inst{\ref{aff55},\ref{aff17}}
\and J.~Gracia-Carpio\inst{\ref{aff56}}
\and A.~Grazian\orcid{0000-0002-5688-0663}\inst{\ref{aff57}}
\and F.~Grupp\inst{\ref{aff56},\ref{aff58}}
\and L.~Guzzo\orcid{0000-0001-8264-5192}\inst{\ref{aff59},\ref{aff1},\ref{aff60}}
\and W.~Holmes\inst{\ref{aff61}}
\and F.~Hormuth\inst{\ref{aff62}}
\and A.~Hornstrup\orcid{0000-0002-3363-0936}\inst{\ref{aff63},\ref{aff64}}
\and K.~Jahnke\orcid{0000-0003-3804-2137}\inst{\ref{aff65}}
\and M.~Jhabvala\inst{\ref{aff66}}
\and B.~Joachimi\orcid{0000-0001-7494-1303}\inst{\ref{aff67}}
\and S.~Kermiche\orcid{0000-0002-0302-5735}\inst{\ref{aff47}}
\and A.~Kiessling\orcid{0000-0002-2590-1273}\inst{\ref{aff61}}
\and B.~Kubik\orcid{0009-0006-5823-4880}\inst{\ref{aff38}}
\and M.~Kunz\orcid{0000-0002-3052-7394}\inst{\ref{aff68}}
\and H.~Kurki-Suonio\orcid{0000-0002-4618-3063}\inst{\ref{aff69},\ref{aff70}}
\and A.~M.~C.~Le~Brun\orcid{0000-0002-0936-4594}\inst{\ref{aff71}}
\and S.~Ligori\orcid{0000-0003-4172-4606}\inst{\ref{aff22}}
\and P.~B.~Lilje\orcid{0000-0003-4324-7794}\inst{\ref{aff72}}
\and V.~Lindholm\orcid{0000-0003-2317-5471}\inst{\ref{aff69},\ref{aff70}}
\and I.~Lloro\orcid{0000-0001-5966-1434}\inst{\ref{aff73}}
\and G.~Mainetti\orcid{0000-0003-2384-2377}\inst{\ref{aff74}}
\and D.~Maino\inst{\ref{aff59},\ref{aff25},\ref{aff60}}
\and E.~Maiorano\orcid{0000-0003-2593-4355}\inst{\ref{aff5}}
\and O.~Mansutti\orcid{0000-0001-5758-4658}\inst{\ref{aff9}}
\and S.~Marcin\inst{\ref{aff75}}
\and O.~Marggraf\orcid{0000-0001-7242-3852}\inst{\ref{aff8}}
\and M.~Martinelli\orcid{0000-0002-6943-7732}\inst{\ref{aff26},\ref{aff27}}
\and N.~Martinet\orcid{0000-0003-2786-7790}\inst{\ref{aff12}}
\and R.~J.~Massey\orcid{0000-0002-6085-3780}\inst{\ref{aff76}}
\and E.~Medinaceli\orcid{0000-0002-4040-7783}\inst{\ref{aff5}}
\and S.~Mei\orcid{0000-0002-2849-559X}\inst{\ref{aff77},\ref{aff78}}
\and M.~Melchior\inst{\ref{aff79}}
\and Y.~Mellier\thanks{Deceased}\inst{\ref{aff80},\ref{aff81}}
\and M.~Meneghetti\orcid{0000-0003-1225-7084}\inst{\ref{aff5},\ref{aff6}}
\and E.~Merlin\orcid{0000-0001-6870-8900}\inst{\ref{aff26}}
\and G.~Meylan\inst{\ref{aff82}}
\and A.~Mora\orcid{0000-0002-1922-8529}\inst{\ref{aff83}}
\and L.~Moscardini\orcid{0000-0002-3473-6716}\inst{\ref{aff4},\ref{aff5},\ref{aff6}}
\and C.~Neissner\orcid{0000-0001-8524-4968}\inst{\ref{aff84},\ref{aff29}}
\and S.-M.~Niemi\orcid{0009-0005-0247-0086}\inst{\ref{aff85}}
\and J.~W.~Nightingale\orcid{0000-0002-8987-7401}\inst{\ref{aff86}}
\and C.~Padilla\orcid{0000-0001-7951-0166}\inst{\ref{aff84}}
\and S.~Paltani\orcid{0000-0002-8108-9179}\inst{\ref{aff45}}
\and F.~Pasian\orcid{0000-0002-4869-3227}\inst{\ref{aff9}}
\and K.~Pedersen\inst{\ref{aff87}}
\and W.~J.~Percival\orcid{0000-0002-0644-5727}\inst{\ref{aff88},\ref{aff89},\ref{aff90}}
\and V.~Pettorino\orcid{0000-0002-4203-9320}\inst{\ref{aff85}}
\and S.~Pires\orcid{0000-0002-0249-2104}\inst{\ref{aff91}}
\and G.~Polenta\orcid{0000-0003-4067-9196}\inst{\ref{aff49}}
\and M.~Poncet\inst{\ref{aff92}}
\and L.~A.~Popa\inst{\ref{aff93}}
\and L.~Pozzetti\orcid{0000-0001-7085-0412}\inst{\ref{aff5}}
\and F.~Raison\orcid{0000-0002-7819-6918}\inst{\ref{aff56}}
\and A.~Renzi\orcid{0000-0001-9856-1970}\inst{\ref{aff94},\ref{aff46},\ref{aff5}}
\and J.~Rhodes\orcid{0000-0002-4485-8549}\inst{\ref{aff61}}
\and G.~Riccio\inst{\ref{aff19}}
\and M.~Roncarelli\orcid{0000-0001-9587-7822}\inst{\ref{aff5}}
\and R.~Saglia\orcid{0000-0003-0378-7032}\inst{\ref{aff58},\ref{aff56}}
\and Z.~Sakr\orcid{0000-0002-4823-3757}\inst{\ref{aff95},\ref{aff96},\ref{aff97}}
\and D.~Sapone\orcid{0000-0001-7089-4503}\inst{\ref{aff98}}
\and B.~Sartoris\orcid{0000-0003-1337-5269}\inst{\ref{aff58},\ref{aff9}}
\and P.~Schneider\orcid{0000-0001-8561-2679}\inst{\ref{aff8}}
\and T.~Schrabback\orcid{0000-0002-6987-7834}\inst{\ref{aff99}}
\and A.~Secroun\orcid{0000-0003-0505-3710}\inst{\ref{aff47}}
\and G.~Seidel\orcid{0000-0003-2907-353X}\inst{\ref{aff65}}
\and S.~Serrano\orcid{0000-0002-0211-2861}\inst{\ref{aff51},\ref{aff100},\ref{aff52}}
\and P.~Simon\inst{\ref{aff8}}
\and C.~Sirignano\orcid{0000-0002-0995-7146}\inst{\ref{aff94},\ref{aff46}}
\and G.~Sirri\orcid{0000-0003-2626-2853}\inst{\ref{aff6}}
\and L.~Stanco\orcid{0000-0002-9706-5104}\inst{\ref{aff46}}
\and J.~Steinwagner\orcid{0000-0001-7443-1047}\inst{\ref{aff56}}
\and P.~Tallada-Cresp\'{i}\orcid{0000-0002-1336-8328}\inst{\ref{aff28},\ref{aff29}}
\and A.~N.~Taylor\inst{\ref{aff23}}
\and I.~Tereno\orcid{0000-0002-4537-6218}\inst{\ref{aff43},\ref{aff101}}
\and N.~Tessore\orcid{0000-0002-9696-7931}\inst{\ref{aff102}}
\and S.~Toft\orcid{0000-0003-3631-7176}\inst{\ref{aff103},\ref{aff104}}
\and R.~Toledo-Moreo\orcid{0000-0002-2997-4859}\inst{\ref{aff105}}
\and F.~Torradeflot\orcid{0000-0003-1160-1517}\inst{\ref{aff29},\ref{aff28}}
\and I.~Tutusaus\orcid{0000-0002-3199-0399}\inst{\ref{aff52},\ref{aff51},\ref{aff96}}
\and L.~Valenziano\orcid{0000-0002-1170-0104}\inst{\ref{aff5},\ref{aff50}}
\and J.~Valiviita\orcid{0000-0001-6225-3693}\inst{\ref{aff69},\ref{aff70}}
\and T.~Vassallo\orcid{0000-0001-6512-6358}\inst{\ref{aff9}}
\and G.~Verdoes~Kleijn\orcid{0000-0001-5803-2580}\inst{\ref{aff106}}
\and Y.~Wang\orcid{0000-0002-4749-2984}\inst{\ref{aff107}}
\and J.~Weller\orcid{0000-0002-8282-2010}\inst{\ref{aff58},\ref{aff56}}
\and G.~Zamorani\orcid{0000-0002-2318-301X}\inst{\ref{aff5}}
\and F.~M.~Zerbi\inst{\ref{aff1}}
\and E.~Zucca\orcid{0000-0002-5845-8132}\inst{\ref{aff5}}
\and V.~Allevato\orcid{0000-0001-7232-5152}\inst{\ref{aff19}}
\and M.~Ballardini\orcid{0000-0003-4481-3559}\inst{\ref{aff108},\ref{aff109},\ref{aff5}}
\and C.~Benoist\inst{\ref{aff110}}
\and M.~Bolzonella\orcid{0000-0003-3278-4607}\inst{\ref{aff5}}
\and E.~Bozzo\orcid{0000-0002-8201-1525}\inst{\ref{aff45}}
\and C.~Burigana\orcid{0000-0002-3005-5796}\inst{\ref{aff111},\ref{aff50}}
\and R.~Cabanac\orcid{0000-0001-6679-2600}\inst{\ref{aff96}}
\and M.~Calabrese\orcid{0000-0002-2637-2422}\inst{\ref{aff112},\ref{aff25}}
\and A.~Cappi\inst{\ref{aff110},\ref{aff5}}
\and T.~Castro\orcid{0000-0002-6292-3228}\inst{\ref{aff9},\ref{aff11},\ref{aff10},\ref{aff15}}
\and J.~A.~Escartin~Vigo\inst{\ref{aff56}}
\and L.~Gabarra\orcid{0000-0002-8486-8856}\inst{\ref{aff113}}
\and J.~Garc\'ia-Bellido\orcid{0000-0002-9370-8360}\inst{\ref{aff114}}
\and V.~Gautard\inst{\ref{aff115}}
\and J.~Macias-Perez\orcid{0000-0002-5385-2763}\inst{\ref{aff116}}
\and R.~Maoli\orcid{0000-0002-6065-3025}\inst{\ref{aff117},\ref{aff26}}
\and J.~Mart\'{i}n-Fleitas\orcid{0000-0002-8594-569X}\inst{\ref{aff118}}
\and N.~Mauri\orcid{0000-0001-8196-1548}\inst{\ref{aff34},\ref{aff6}}
\and R.~B.~Metcalf\orcid{0000-0003-3167-2574}\inst{\ref{aff4},\ref{aff5}}
\and P.~Monaco\orcid{0000-0003-2083-7564}\inst{\ref{aff119},\ref{aff9},\ref{aff11},\ref{aff10}}
\and A.~Pezzotta\orcid{0000-0003-0726-2268}\inst{\ref{aff1}}
\and M.~P\"ontinen\orcid{0000-0001-5442-2530}\inst{\ref{aff69}}
\and I.~Risso\orcid{0000-0003-2525-7761}\inst{\ref{aff1},\ref{aff2}}
\and V.~Scottez\orcid{0009-0008-3864-940X}\inst{\ref{aff80},\ref{aff120}}
\and M.~Sereno\orcid{0000-0003-0302-0325}\inst{\ref{aff5},\ref{aff6}}
\and M.~Tenti\orcid{0000-0002-4254-5901}\inst{\ref{aff6}}
\and M.~Tucci\inst{\ref{aff45}}
\and M.~Wiesmann\orcid{0009-0000-8199-5860}\inst{\ref{aff72}}
\and Y.~Akrami\orcid{0000-0002-2407-7956}\inst{\ref{aff114},\ref{aff121}}
\and G.~Alguero\inst{\ref{aff116}}
\and I.~T.~Andika\orcid{0000-0001-6102-9526}\inst{\ref{aff54}}
\and G.~Angora\orcid{0000-0002-0316-6562}\inst{\ref{aff19},\ref{aff108}}
\and S.~Anselmi\orcid{0000-0002-3579-9583}\inst{\ref{aff46},\ref{aff94},\ref{aff122}}
\and M.~Archidiacono\orcid{0000-0003-4952-9012}\inst{\ref{aff59},\ref{aff60}}
\and F.~Atrio-Barandela\orcid{0000-0002-2130-2513}\inst{\ref{aff123}}
\and E.~Aubourg\orcid{0000-0002-5592-023X}\inst{\ref{aff77},\ref{aff124}}
\and L.~Bazzanini\orcid{0000-0003-0727-0137}\inst{\ref{aff108},\ref{aff5}}
\and J.~Bel\inst{\ref{aff125}}
\and D.~Bertacca\orcid{0000-0002-2490-7139}\inst{\ref{aff94},\ref{aff57},\ref{aff46}}
\and M.~Bethermin\orcid{0000-0002-3915-2015}\inst{\ref{aff126}}
\and F.~Beutler\orcid{0000-0003-0467-5438}\inst{\ref{aff23}}
\and A.~Blanchard\orcid{0000-0001-8555-9003}\inst{\ref{aff96}}
\and L.~Blot\orcid{0000-0002-9622-7167}\inst{\ref{aff127},\ref{aff71}}
\and H.~B\"ohringer\orcid{0000-0001-8241-4204}\inst{\ref{aff56},\ref{aff54},\ref{aff128}}
\and M.~Bonici\orcid{0000-0002-8430-126X}\inst{\ref{aff88},\ref{aff25}}
\and S.~Borgani\orcid{0000-0001-6151-6439}\inst{\ref{aff119},\ref{aff10},\ref{aff9},\ref{aff11},\ref{aff15}}
\and M.~L.~Brown\orcid{0000-0002-0370-8077}\inst{\ref{aff36}}
\and S.~Bruton\orcid{0000-0002-6503-5218}\inst{\ref{aff129}}
\and A.~Calabro\orcid{0000-0003-2536-1614}\inst{\ref{aff26}}
\and B.~Camacho~Quevedo\orcid{0000-0002-8789-4232}\inst{\ref{aff10},\ref{aff14},\ref{aff9}}
\and F.~Caro\inst{\ref{aff26}}
\and C.~S.~Carvalho\inst{\ref{aff101}}
\and F.~Cogato\orcid{0000-0003-4632-6113}\inst{\ref{aff4},\ref{aff5}}
\and S.~Conseil\orcid{0000-0002-3657-4191}\inst{\ref{aff38}}
\and A.~R.~Cooray\orcid{0000-0002-3892-0190}\inst{\ref{aff130}}
\and O.~Cucciati\orcid{0000-0002-9336-7551}\inst{\ref{aff5}}
\and S.~Davini\orcid{0000-0003-3269-1718}\inst{\ref{aff2}}
\and G.~Desprez\orcid{0000-0001-8325-1742}\inst{\ref{aff106}}
\and A.~D\'iaz-S\'anchez\orcid{0000-0003-0748-4768}\inst{\ref{aff131}}
\and S.~Di~Domizio\orcid{0000-0003-2863-5895}\inst{\ref{aff3},\ref{aff2}}
\and J.~M.~Diego\orcid{0000-0001-9065-3926}\inst{\ref{aff132}}
\and V.~Duret\orcid{0009-0009-0383-4960}\inst{\ref{aff47}}
\and M.~Y.~Elkhashab\orcid{0000-0001-9306-2603}\inst{\ref{aff9},\ref{aff11},\ref{aff119},\ref{aff10}}
\and A.~Enia\orcid{0000-0002-0200-2857}\inst{\ref{aff5}}
\and Y.~Fang\orcid{0000-0002-0334-6950}\inst{\ref{aff58}}
\and A.~G.~Ferrari\orcid{0009-0005-5266-4110}\inst{\ref{aff6}}
\and A.~Finoguenov\orcid{0000-0002-4606-5403}\inst{\ref{aff69}}
\and F.~Fontanot\orcid{0000-0003-4744-0188}\inst{\ref{aff9},\ref{aff10}}
\and A.~Franco\orcid{0000-0002-4761-366X}\inst{\ref{aff133},\ref{aff134},\ref{aff135}}
\and K.~Ganga\orcid{0000-0001-8159-8208}\inst{\ref{aff77}}
\and T.~Gasparetto\orcid{0000-0002-7913-4866}\inst{\ref{aff26}}
\and E.~Gaztanaga\orcid{0000-0001-9632-0815}\inst{\ref{aff52},\ref{aff51},\ref{aff136}}
\and F.~Giacomini\orcid{0000-0002-3129-2814}\inst{\ref{aff6}}
\and F.~Gianotti\orcid{0000-0003-4666-119X}\inst{\ref{aff5}}
\and G.~Gozaliasl\orcid{0000-0002-0236-919X}\inst{\ref{aff137},\ref{aff69}}
\and A.~Gruppuso\orcid{0000-0001-9272-5292}\inst{\ref{aff5},\ref{aff6}}
\and C.~M.~Gutierrez\orcid{0000-0001-7854-783X}\inst{\ref{aff35},\ref{aff138}}
\and A.~Hall\orcid{0000-0002-3139-8651}\inst{\ref{aff23}}
\and H.~Hildebrandt\orcid{0000-0002-9814-3338}\inst{\ref{aff139}}
\and J.~Hjorth\orcid{0000-0002-4571-2306}\inst{\ref{aff87}}
\and S.~Joudaki\orcid{0000-0001-8820-673X}\inst{\ref{aff28},\ref{aff136}}
\and J.~J.~E.~Kajava\orcid{0000-0002-3010-8333}\inst{\ref{aff140},\ref{aff141},\ref{aff142}}
\and Y.~Kang\orcid{0009-0000-8588-7250}\inst{\ref{aff45}}
\and V.~Kansal\orcid{0000-0002-4008-6078}\inst{\ref{aff143},\ref{aff144}}
\and D.~Karagiannis\orcid{0000-0002-4927-0816}\inst{\ref{aff108},\ref{aff145}}
\and K.~Kiiveri\inst{\ref{aff13}}
\and J.~Kim\orcid{0000-0003-2776-2761}\inst{\ref{aff113}}
\and C.~C.~Kirkpatrick\inst{\ref{aff13}}
\and S.~Kruk\orcid{0000-0001-8010-8879}\inst{\ref{aff17}}
\and M.~Lattanzi\orcid{0000-0003-1059-2532}\inst{\ref{aff109}}
\and L.~Legrand\orcid{0000-0003-0610-5252}\inst{\ref{aff146},\ref{aff147}}
\and M.~Lembo\orcid{0000-0002-5271-5070}\inst{\ref{aff81}}
\and F.~Lepori\orcid{0009-0000-5061-7138}\inst{\ref{aff148}}
\and G.~Leroy\orcid{0009-0004-2523-4425}\inst{\ref{aff149},\ref{aff76}}
\and G.~F.~Lesci\orcid{0000-0002-4607-2830}\inst{\ref{aff4},\ref{aff5}}
\and J.~Lesgourgues\orcid{0000-0001-7627-353X}\inst{\ref{aff30}}
\and T.~I.~Liaudat\orcid{0000-0002-9104-314X}\inst{\ref{aff124}}
\and S.~J.~Liu\orcid{0000-0001-7680-2139}\inst{\ref{aff48}}
\and A.~Loureiro\orcid{0000-0002-4371-0876}\inst{\ref{aff150},\ref{aff151}}
\and M.~Magliocchetti\orcid{0000-0001-9158-4838}\inst{\ref{aff48}}
\and F.~Mannucci\orcid{0000-0002-4803-2381}\inst{\ref{aff152}}
\and C.~J.~A.~P.~Martins\orcid{0000-0002-4886-9261}\inst{\ref{aff153},\ref{aff154}}
\and L.~Maurin\orcid{0000-0002-8406-0857}\inst{\ref{aff16}}
\and M.~Migliaccio\inst{\ref{aff155},\ref{aff156}}
\and M.~Miluzio\inst{\ref{aff17},\ref{aff157}}
\and C.~Moretti\orcid{0000-0003-3314-8936}\inst{\ref{aff9},\ref{aff10},\ref{aff11}}
\and G.~Morgante\inst{\ref{aff5}}
\and S.~Nadathur\orcid{0000-0001-9070-3102}\inst{\ref{aff136}}
\and K.~Naidoo\orcid{0000-0002-9182-1802}\inst{\ref{aff136},\ref{aff65}}
\and P.~Natoli\orcid{0000-0003-0126-9100}\inst{\ref{aff108},\ref{aff109}}
\and A.~Navarro-Alsina\orcid{0000-0002-3173-2592}\inst{\ref{aff8}}
\and S.~Nesseris\orcid{0000-0002-0567-0324}\inst{\ref{aff114}}
\and L.~Pagano\orcid{0000-0003-1820-5998}\inst{\ref{aff108},\ref{aff109}}
\and D.~Paoletti\orcid{0000-0003-4761-6147}\inst{\ref{aff5},\ref{aff50}}
\and F.~Passalacqua\orcid{0000-0002-8606-4093}\inst{\ref{aff94},\ref{aff46}}
\and K.~Paterson\orcid{0000-0001-8340-3486}\inst{\ref{aff65}}
\and L.~Patrizii\inst{\ref{aff6}}
\and R.~Paviot\orcid{0009-0002-8108-3460}\inst{\ref{aff91}}
\and A.~Pisani\orcid{0000-0002-6146-4437}\inst{\ref{aff47}}
\and D.~Potter\orcid{0000-0002-0757-5195}\inst{\ref{aff148}}
\and G.~W.~Pratt\inst{\ref{aff91}}
\and S.~Quai\orcid{0000-0002-0449-8163}\inst{\ref{aff4},\ref{aff5}}
\and M.~Radovich\orcid{0000-0002-3585-866X}\inst{\ref{aff57}}
\and K.~Rojas\orcid{0000-0003-1391-6854}\inst{\ref{aff75}}
\and W.~Roster\orcid{0000-0002-9149-6528}\inst{\ref{aff56}}
\and S.~Sacquegna\orcid{0000-0002-8433-6630}\inst{\ref{aff158}}
\and M.~Sahl\'en\orcid{0000-0003-0973-4804}\inst{\ref{aff159}}
\and D.~B.~Sanders\orcid{0000-0002-1233-9998}\inst{\ref{aff33}}
\and E.~Sarpa\orcid{0000-0002-1256-655X}\inst{\ref{aff14},\ref{aff15},\ref{aff9}}
\and A.~Schneider\orcid{0000-0001-7055-8104}\inst{\ref{aff148}}
\and D.~Sciotti\orcid{0009-0008-4519-2620}\inst{\ref{aff26},\ref{aff27}}
\and E.~Sellentin\inst{\ref{aff160},\ref{aff24}}
\and L.~C.~Smith\orcid{0000-0002-3259-2771}\inst{\ref{aff161}}
\and J.~G.~Sorce\orcid{0000-0002-2307-2432}\inst{\ref{aff162},\ref{aff16}}
\and K.~Tanidis\orcid{0000-0001-9843-5130}\inst{\ref{aff113}}
\and C.~Tao\orcid{0000-0001-7961-8177}\inst{\ref{aff47}}
\and F.~Tarsitano\orcid{0000-0002-5919-0238}\inst{\ref{aff163},\ref{aff45}}
\and G.~Testera\inst{\ref{aff2}}
\and R.~Teyssier\orcid{0000-0001-7689-0933}\inst{\ref{aff164}}
\and S.~Tosi\orcid{0000-0002-7275-9193}\inst{\ref{aff3},\ref{aff2},\ref{aff1}}
\and A.~Troja\orcid{0000-0003-0239-4595}\inst{\ref{aff94},\ref{aff46}}
\and A.~Venhola\orcid{0000-0001-6071-4564}\inst{\ref{aff165}}
\and D.~Vergani\orcid{0000-0003-0898-2216}\inst{\ref{aff5}}
\and F.~Vernizzi\orcid{0000-0003-3426-2802}\inst{\ref{aff166}}
\and G.~Verza\orcid{0000-0002-1886-8348}\inst{\ref{aff167},\ref{aff168}}
\and P.~Vielzeuf\orcid{0000-0003-2035-9339}\inst{\ref{aff47}}
\and S.~Vinciguerra\orcid{0009-0005-4018-3184}\inst{\ref{aff12}}
\and N.~A.~Walton\orcid{0000-0003-3983-8778}\inst{\ref{aff161}}
\and A.~H.~Wright\orcid{0000-0001-7363-7932}\inst{\ref{aff139}}}
										   
\institute{INAF-Osservatorio Astronomico di Brera, Via Brera 28, 20122 Milano, Italy\label{aff1}
\and
INFN-Sezione di Genova, Via Dodecaneso 33, 16146, Genova, Italy\label{aff2}
\and
Dipartimento di Fisica, Universit\`a di Genova, Via Dodecaneso 33, 16146, Genova, Italy\label{aff3}
\and
Dipartimento di Fisica e Astronomia "Augusto Righi" - Alma Mater Studiorum Universit\`a di Bologna, via Piero Gobetti 93/2, 40129 Bologna, Italy\label{aff4}
\and
INAF-Osservatorio di Astrofisica e Scienza dello Spazio di Bologna, Via Piero Gobetti 93/3, 40129 Bologna, Italy\label{aff5}
\and
INFN-Sezione di Bologna, Viale Berti Pichat 6/2, 40127 Bologna, Italy\label{aff6}
\and
Dipartimento di Fisica e Astronomia, Universit\`a di Bologna, Via Gobetti 93/2, 40129 Bologna, Italy\label{aff7}
\and
Universit\"at Bonn, Argelander-Institut f\"ur Astronomie, Auf dem H\"ugel 71, 53121 Bonn, Germany\label{aff8}
\and
INAF-Osservatorio Astronomico di Trieste, Via G. B. Tiepolo 11, 34143 Trieste, Italy\label{aff9}
\and
IFPU, Institute for Fundamental Physics of the Universe, via Beirut 2, 34151 Trieste, Italy\label{aff10}
\and
INFN, Sezione di Trieste, Via Valerio 2, 34127 Trieste TS, Italy\label{aff11}
\and
Aix-Marseille Universit\'e, CNRS, CNES, LAM, Marseille, France\label{aff12}
\and
Department of Physics and Helsinki Institute of Physics, Gustaf H\"allstr\"omin katu 2, University of Helsinki, 00014 Helsinki, Finland\label{aff13}
\and
SISSA, International School for Advanced Studies, Via Bonomea 265, 34136 Trieste TS, Italy\label{aff14}
\and
ICSC - Centro Nazionale di Ricerca in High Performance Computing, Big Data e Quantum Computing, Via Magnanelli 2, Bologna, Italy\label{aff15}
\and
Universit\'e Paris-Saclay, CNRS, Institut d'astrophysique spatiale, 91405, Orsay, France\label{aff16}
\and
ESAC/ESA, Camino Bajo del Castillo, s/n., Urb. Villafranca del Castillo, 28692 Villanueva de la Ca\~nada, Madrid, Spain\label{aff17}
\and
Department of Physics "E. Pancini", University Federico II, Via Cinthia 6, 80126, Napoli, Italy\label{aff18}
\and
INAF-Osservatorio Astronomico di Capodimonte, Via Moiariello 16, 80131 Napoli, Italy\label{aff19}
\and
Dipartimento di Fisica, Universit\`a degli Studi di Torino, Via P. Giuria 1, 10125 Torino, Italy\label{aff20}
\and
INFN-Sezione di Torino, Via P. Giuria 1, 10125 Torino, Italy\label{aff21}
\and
INAF-Osservatorio Astrofisico di Torino, Via Osservatorio 20, 10025 Pino Torinese (TO), Italy\label{aff22}
\and
Institute for Astronomy, University of Edinburgh, Royal Observatory, Blackford Hill, Edinburgh EH9 3HJ, UK\label{aff23}
\and
Leiden Observatory, Leiden University, Einsteinweg 55, 2333 CC Leiden, The Netherlands\label{aff24}
\and
INAF-IASF Milano, Via Alfonso Corti 12, 20133 Milano, Italy\label{aff25}
\and
INAF-Osservatorio Astronomico di Roma, Via Frascati 33, 00078 Monteporzio Catone, Italy\label{aff26}
\and
INFN-Sezione di Roma, Piazzale Aldo Moro, 2 - c/o Dipartimento di Fisica, Edificio G. Marconi, 00185 Roma, Italy\label{aff27}
\and
Centro de Investigaciones Energ\'eticas, Medioambientales y Tecnol\'ogicas (CIEMAT), Avenida Complutense 40, 28040 Madrid, Spain\label{aff28}
\and
Port d'Informaci\'{o} Cient\'{i}fica, Campus UAB, C. Albareda s/n, 08193 Bellaterra (Barcelona), Spain\label{aff29}
\and
Institute for Theoretical Particle Physics and Cosmology (TTK), RWTH Aachen University, 52056 Aachen, Germany\label{aff30}
\and
Deutsches Zentrum f\"ur Luft- und Raumfahrt e. V. (DLR), Linder H\"ohe, 51147 K\"oln, Germany\label{aff31}
\and
INFN section of Naples, Via Cinthia 6, 80126, Napoli, Italy\label{aff32}
\and
Institute for Astronomy, University of Hawaii, 2680 Woodlawn Drive, Honolulu, HI 96822, USA\label{aff33}
\and
Dipartimento di Fisica e Astronomia "Augusto Righi" - Alma Mater Studiorum Universit\`a di Bologna, Viale Berti Pichat 6/2, 40127 Bologna, Italy\label{aff34}
\and
Instituto de Astrof\'{\i}sica de Canarias, E-38205 La Laguna, Tenerife, Spain\label{aff35}
\and
Jodrell Bank Centre for Astrophysics, Department of Physics and Astronomy, University of Manchester, Oxford Road, Manchester M13 9PL, UK\label{aff36}
\and
European Space Agency/ESRIN, Largo Galileo Galilei 1, 00044 Frascati, Roma, Italy\label{aff37}
\and
Universit\'e Claude Bernard Lyon 1, CNRS/IN2P3, IP2I Lyon, UMR 5822, Villeurbanne, F-69100, France\label{aff38}
\and
Institut de Ci\`{e}ncies del Cosmos (ICCUB), Universitat de Barcelona (IEEC-UB), Mart\'{i} i Franqu\`{e}s 1, 08028 Barcelona, Spain\label{aff39}
\and
Instituci\'o Catalana de Recerca i Estudis Avan\c{c}ats (ICREA), Passeig de Llu\'{\i}s Companys 23, 08010 Barcelona, Spain\label{aff40}
\and
Institut de Ciencies de l'Espai (IEEC-CSIC), Campus UAB, Carrer de Can Magrans, s/n Cerdanyola del Vall\'es, 08193 Barcelona, Spain\label{aff41}
\and
UCB Lyon 1, CNRS/IN2P3, IUF, IP2I Lyon, 4 rue Enrico Fermi, 69622 Villeurbanne, France\label{aff42}
\and
Departamento de F\'isica, Faculdade de Ci\^encias, Universidade de Lisboa, Edif\'icio C8, Campo Grande, PT1749-016 Lisboa, Portugal\label{aff43}
\and
Instituto de Astrof\'isica e Ci\^encias do Espa\c{c}o, Faculdade de Ci\^encias, Universidade de Lisboa, Campo Grande, 1749-016 Lisboa, Portugal\label{aff44}
\and
Department of Astronomy, University of Geneva, ch. d'Ecogia 16, 1290 Versoix, Switzerland\label{aff45}
\and
INFN-Padova, Via Marzolo 8, 35131 Padova, Italy\label{aff46}
\and
Aix-Marseille Universit\'e, CNRS/IN2P3, CPPM, Marseille, France\label{aff47}
\and
INAF-Istituto di Astrofisica e Planetologia Spaziali, via del Fosso del Cavaliere, 100, 00100 Roma, Italy\label{aff48}
\and
Space Science Data Center, Italian Space Agency, via del Politecnico snc, 00133 Roma, Italy\label{aff49}
\and
INFN-Bologna, Via Irnerio 46, 40126 Bologna, Italy\label{aff50}
\and
Institut d'Estudis Espacials de Catalunya (IEEC),  Edifici RDIT, Campus UPC, 08860 Castelldefels, Barcelona, Spain\label{aff51}
\and
Institute of Space Sciences (ICE, CSIC), Campus UAB, Carrer de Can Magrans, s/n, 08193 Barcelona, Spain\label{aff52}
\and
School of Physics, HH Wills Physics Laboratory, University of Bristol, Tyndall Avenue, Bristol, BS8 1TL, UK\label{aff53}
\and
University Observatory, LMU Faculty of Physics, Scheinerstr.~1, 81679 Munich, Germany\label{aff54}
\and
FRACTAL S.L.N.E., calle Tulip\'an 2, Portal 13 1A, 28231, Las Rozas de Madrid, Spain\label{aff55}
\and
Max Planck Institute for Extraterrestrial Physics, Giessenbachstr. 1, 85748 Garching, Germany\label{aff56}
\and
INAF-Osservatorio Astronomico di Padova, Via dell'Osservatorio 5, 35122 Padova, Italy\label{aff57}
\and
Universit\"ats-Sternwarte M\"unchen, Fakult\"at f\"ur Physik, Ludwig-Maximilians-Universit\"at M\"unchen, Scheinerstr.~1, 81679 M\"unchen, Germany\label{aff58}
\and
Dipartimento di Fisica "Aldo Pontremoli", Universit\`a degli Studi di Milano, Via Celoria 16, 20133 Milano, Italy\label{aff59}
\and
INFN-Sezione di Milano, Via Celoria 16, 20133 Milano, Italy\label{aff60}
\and
Jet Propulsion Laboratory, California Institute of Technology, 4800 Oak Grove Drive, Pasadena, CA, 91109, USA\label{aff61}
\and
Felix Hormuth Engineering, Goethestr. 17, 69181 Leimen, Germany\label{aff62}
\and
Technical University of Denmark, Elektrovej 327, 2800 Kgs. Lyngby, Denmark\label{aff63}
\and
Cosmic Dawn Center (DAWN), Denmark\label{aff64}
\and
Max-Planck-Institut f\"ur Astronomie, K\"onigstuhl 17, 69117 Heidelberg, Germany\label{aff65}
\and
NASA Goddard Space Flight Center, Greenbelt, MD 20771, USA\label{aff66}
\and
Department of Physics and Astronomy, University College London, Gower Street, London WC1E 6BT, UK\label{aff67}
\and
Universit\'e de Gen\`eve, D\'epartement de Physique Th\'eorique and Centre for Astroparticle Physics, 24 quai Ernest-Ansermet, CH-1211 Gen\`eve 4, Switzerland\label{aff68}
\and
Department of Physics, P.O. Box 64, University of Helsinki, 00014 Helsinki, Finland\label{aff69}
\and
Helsinki Institute of Physics, Gustaf H{\"a}llstr{\"o}min katu 2, University of Helsinki, 00014 Helsinki, Finland\label{aff70}
\and
Laboratoire d'etude de l'Univers et des phenomenes eXtremes, Observatoire de Paris, Universit\'e PSL, Sorbonne Universit\'e, CNRS, 92190 Meudon, France\label{aff71}
\and
Institute of Theoretical Astrophysics, University of Oslo, P.O. Box 1029 Blindern, 0315 Oslo, Norway\label{aff72}
\and
SKAO, Jodrell Bank, Lower Withington, Macclesfield SK11 9FT, UK\label{aff73}
\and
Centre de Calcul de l'IN2P3/CNRS, 21 avenue Pierre de Coubertin 69627 Villeurbanne Cedex, France\label{aff74}
\and
University of Applied Sciences and Arts of Northwestern Switzerland, School of Computer Science, 5210 Windisch, Switzerland\label{aff75}
\and
Department of Physics, Institute for Computational Cosmology, Durham University, South Road, Durham, DH1 3LE, UK\label{aff76}
\and
Universit\'e Paris Cit\'e, CNRS, Astroparticule et Cosmologie, 75013 Paris, France\label{aff77}
\and
CNRS-UCB International Research Laboratory, Centre Pierre Bin\'etruy, IRL2007, CPB-IN2P3, Berkeley, USA\label{aff78}
\and
University of Applied Sciences and Arts of Northwestern Switzerland, School of Engineering, 5210 Windisch, Switzerland\label{aff79}
\and
Institut d'Astrophysique de Paris, 98bis Boulevard Arago, 75014, Paris, France\label{aff80}
\and
Institut d'Astrophysique de Paris, UMR 7095, CNRS, and Sorbonne Universit\'e, 98 bis boulevard Arago, 75014 Paris, France\label{aff81}
\and
Institute of Physics, Laboratory of Astrophysics, Ecole Polytechnique F\'ed\'erale de Lausanne (EPFL), Observatoire de Sauverny, 1290 Versoix, Switzerland\label{aff82}
\and
Telespazio UK S.L. for European Space Agency (ESA), Camino bajo del Castillo, s/n, Urbanizacion Villafranca del Castillo, Villanueva de la Ca\~nada, 28692 Madrid, Spain\label{aff83}
\and
Institut de F\'{i}sica d'Altes Energies (IFAE), The Barcelona Institute of Science and Technology, Campus UAB, 08193 Bellaterra (Barcelona), Spain\label{aff84}
\and
European Space Agency/ESTEC, Keplerlaan 1, 2201 AZ Noordwijk, The Netherlands\label{aff85}
\and
School of Mathematics, Statistics and Physics, Newcastle University, Herschel Building, Newcastle-upon-Tyne, NE1 7RU, UK\label{aff86}
\and
DARK, Niels Bohr Institute, University of Copenhagen, Jagtvej 155, 2200 Copenhagen, Denmark\label{aff87}
\and
Waterloo Centre for Astrophysics, University of Waterloo, Waterloo, Ontario N2L 3G1, Canada\label{aff88}
\and
Department of Physics and Astronomy, University of Waterloo, Waterloo, Ontario N2L 3G1, Canada\label{aff89}
\and
Perimeter Institute for Theoretical Physics, Waterloo, Ontario N2L 2Y5, Canada\label{aff90}
\and
Universit\'e Paris-Saclay, Universit\'e Paris Cit\'e, CEA, CNRS, AIM, 91191, Gif-sur-Yvette, France\label{aff91}
\and
Centre National d'Etudes Spatiales -- Centre spatial de Toulouse, 18 avenue Edouard Belin, 31401 Toulouse Cedex 9, France\label{aff92}
\and
Institute of Space Science, Str. Atomistilor, nr. 409 M\u{a}gurele, Ilfov, 077125, Romania\label{aff93}
\and
Dipartimento di Fisica e Astronomia "G. Galilei", Universit\`a di Padova, Via Marzolo 8, 35131 Padova, Italy\label{aff94}
\and
Institut f\"ur Theoretische Physik, University of Heidelberg, Philosophenweg 16, 69120 Heidelberg, Germany\label{aff95}
\and
Institut de Recherche en Astrophysique et Plan\'etologie (IRAP), Universit\'e de Toulouse, CNRS, UPS, CNES, 14 Av. Edouard Belin, 31400 Toulouse, France\label{aff96}
\and
Universit\'e St Joseph; Faculty of Sciences, Beirut, Lebanon\label{aff97}
\and
Departamento de F\'isica, FCFM, Universidad de Chile, Blanco Encalada 2008, Santiago, Chile\label{aff98}
\and
Universit\"at Innsbruck, Institut f\"ur Astro- und Teilchenphysik, Technikerstr. 25/8, 6020 Innsbruck, Austria\label{aff99}
\and
Satlantis, University Science Park, Sede Bld 48940, Leioa-Bilbao, Spain\label{aff100}
\and
Instituto de Astrof\'isica e Ci\^encias do Espa\c{c}o, Faculdade de Ci\^encias, Universidade de Lisboa, Tapada da Ajuda, 1349-018 Lisboa, Portugal\label{aff101}
\and
Mullard Space Science Laboratory, University College London, Holmbury St Mary, Dorking, Surrey RH5 6NT, UK\label{aff102}
\and
Cosmic Dawn Center (DAWN)\label{aff103}
\and
Niels Bohr Institute, University of Copenhagen, Jagtvej 128, 2200 Copenhagen, Denmark\label{aff104}
\and
Universidad Polit\'ecnica de Cartagena, Departamento de Electr\'onica y Tecnolog\'ia de Computadoras,  Plaza del Hospital 1, 30202 Cartagena, Spain\label{aff105}
\and
Kapteyn Astronomical Institute, University of Groningen, PO Box 800, 9700 AV Groningen, The Netherlands\label{aff106}
\and
Caltech/IPAC, 1200 E. California Blvd., Pasadena, CA 91125, USA\label{aff107}
\and
Dipartimento di Fisica e Scienze della Terra, Universit\`a degli Studi di Ferrara, Via Giuseppe Saragat 1, 44122 Ferrara, Italy\label{aff108}
\and
Istituto Nazionale di Fisica Nucleare, Sezione di Ferrara, Via Giuseppe Saragat 1, 44122 Ferrara, Italy\label{aff109}
\and
Universit\'e C\^{o}te d'Azur, Observatoire de la C\^{o}te d'Azur, CNRS, Laboratoire Lagrange, Bd de l'Observatoire, CS 34229, 06304 Nice cedex 4, France\label{aff110}
\and
INAF, Istituto di Radioastronomia, Via Piero Gobetti 101, 40129 Bologna, Italy\label{aff111}
\and
Astronomical Observatory of the Autonomous Region of the Aosta Valley (OAVdA), Loc. Lignan 39, I-11020, Nus (Aosta Valley), Italy\label{aff112}
\and
Department of Physics, Oxford University, Keble Road, Oxford OX1 3RH, UK\label{aff113}
\and
Instituto de F\'isica Te\'orica UAM-CSIC, Campus de Cantoblanco, 28049 Madrid, Spain\label{aff114}
\and
CEA Saclay, DFR/IRFU, Service d'Astrophysique, Bat. 709, 91191 Gif-sur-Yvette, France\label{aff115}
\and
Univ. Grenoble Alpes, CNRS, Grenoble INP, LPSC-IN2P3, 53, Avenue des Martyrs, 38000, Grenoble, France\label{aff116}
\and
Dipartimento di Fisica, Sapienza Universit\`a di Roma, Piazzale Aldo Moro 2, 00185 Roma, Italy\label{aff117}
\and
Aurora Technology for European Space Agency (ESA), Camino bajo del Castillo, s/n, Urbanizacion Villafranca del Castillo, Villanueva de la Ca\~nada, 28692 Madrid, Spain\label{aff118}
\and
Dipartimento di Fisica - Sezione di Astronomia, Universit\`a di Trieste, Via Tiepolo 11, 34131 Trieste, Italy\label{aff119}
\and
ICL, Junia, Universit\'e Catholique de Lille, LITL, 59000 Lille, France\label{aff120}
\and
CERCA/ISO, Department of Physics, Case Western Reserve University, 10900 Euclid Avenue, Cleveland, OH 44106, USA\label{aff121}
\and
Laboratoire Univers et Th\'eorie, Observatoire de Paris, Universit\'e PSL, Universit\'e Paris Cit\'e, CNRS, 92190 Meudon, France\label{aff122}
\and
Departamento de F{\'\i}sica Fundamental. Universidad de Salamanca. Plaza de la Merced s/n. 37008 Salamanca, Spain\label{aff123}
\and
IRFU, CEA, Universit\'e Paris-Saclay 91191 Gif-sur-Yvette Cedex, France\label{aff124}
\and
Aix-Marseille Universit\'e, Universit\'e de Toulon, CNRS, CPT, Marseille, France\label{aff125}
\and
Universit\'e de Strasbourg, CNRS, Observatoire astronomique de Strasbourg, UMR 7550, 67000 Strasbourg, France\label{aff126}
\and
Center for Data-Driven Discovery, Kavli IPMU (WPI), UTIAS, The University of Tokyo, Kashiwa, Chiba 277-8583, Japan\label{aff127}
\and
Max-Planck-Institut f\"ur Physik, Boltzmannstr. 8, 85748 Garching, Germany\label{aff128}
\and
California Institute of Technology, 1200 E California Blvd, Pasadena, CA 91125, USA\label{aff129}
\and
Department of Physics \& Astronomy, University of California Irvine, Irvine CA 92697, USA\label{aff130}
\and
Departamento F\'isica Aplicada, Universidad Polit\'ecnica de Cartagena, Campus Muralla del Mar, 30202 Cartagena, Murcia, Spain\label{aff131}
\and
Instituto de F\'isica de Cantabria, Edificio Juan Jord\'a, Avenida de los Castros, 39005 Santander, Spain\label{aff132}
\and
INFN, Sezione di Lecce, Via per Arnesano, CP-193, 73100, Lecce, Italy\label{aff133}
\and
Department of Mathematics and Physics E. De Giorgi, University of Salento, Via per Arnesano, CP-I93, 73100, Lecce, Italy\label{aff134}
\and
INAF-Sezione di Lecce, c/o Dipartimento Matematica e Fisica, Via per Arnesano, 73100, Lecce, Italy\label{aff135}
\and
Institute of Cosmology and Gravitation, University of Portsmouth, Portsmouth PO1 3FX, UK\label{aff136}
\and
Department of Computer Science, Aalto University, PO Box 15400, Espoo, FI-00 076, Finland\label{aff137}
\and
Universidad de La Laguna, Dpto. Astrof\'\i sica, E-38206 La Laguna, Tenerife, Spain\label{aff138}
\and
Ruhr University Bochum, Faculty of Physics and Astronomy, Astronomical Institute (AIRUB), German Centre for Cosmological Lensing (GCCL), 44780 Bochum, Germany\label{aff139}
\and
Department of Physics and Astronomy, Vesilinnantie 5, University of Turku, 20014 Turku, Finland\label{aff140}
\and
Finnish Centre for Astronomy with ESO (FINCA), Quantum, Vesilinnantie 5, University of Turku, 20014 Turku, Finland\label{aff141}
\and
Serco for European Space Agency (ESA), Camino bajo del Castillo, s/n, Urbanizacion Villafranca del Castillo, Villanueva de la Ca\~nada, 28692 Madrid, Spain\label{aff142}
\and
ARC Centre of Excellence for Dark Matter Particle Physics, Melbourne, Australia\label{aff143}
\and
Centre for Astrophysics \& Supercomputing, Swinburne University of Technology,  Hawthorn, Victoria 3122, Australia\label{aff144}
\and
Department of Physics and Astronomy, University of the Western Cape, Bellville, Cape Town, 7535, South Africa\label{aff145}
\and
DAMTP, Centre for Mathematical Sciences, Wilberforce Road, Cambridge CB3 0WA, UK\label{aff146}
\and
Kavli Institute for Cosmology Cambridge, Madingley Road, Cambridge, CB3 0HA, UK\label{aff147}
\and
Department of Astrophysics, University of Zurich, Winterthurerstrasse 190, 8057 Zurich, Switzerland\label{aff148}
\and
Department of Physics, Centre for Extragalactic Astronomy, Durham University, South Road, Durham, DH1 3LE, UK\label{aff149}
\and
Oskar Klein Centre for Cosmoparticle Physics, Department of Physics, Stockholm University, Stockholm, SE-106 91, Sweden\label{aff150}
\and
Astrophysics Group, Blackett Laboratory, Imperial College London, London SW7 2AZ, UK\label{aff151}
\and
INAF-Osservatorio Astrofisico di Arcetri, Largo E. Fermi 5, 50125, Firenze, Italy\label{aff152}
\and
Centro de Astrof\'{\i}sica da Universidade do Porto, Rua das Estrelas, 4150-762 Porto, Portugal\label{aff153}
\and
Instituto de Astrof\'isica e Ci\^encias do Espa\c{c}o, Universidade do Porto, CAUP, Rua das Estrelas, PT4150-762 Porto, Portugal\label{aff154}
\and
Dipartimento di Fisica, Universit\`a di Roma Tor Vergata, Via della Ricerca Scientifica 1, Roma, Italy\label{aff155}
\and
INFN, Sezione di Roma 2, Via della Ricerca Scientifica 1, Roma, Italy\label{aff156}
\and
HE Space for European Space Agency (ESA), Camino bajo del Castillo, s/n, Urbanizacion Villafranca del Castillo, Villanueva de la Ca\~nada, 28692 Madrid, Spain\label{aff157}
\and
INAF - Osservatorio Astronomico d'Abruzzo, Via Maggini, 64100, Teramo, Italy\label{aff158}
\and
Theoretical astrophysics, Department of Physics and Astronomy, Uppsala University, Box 516, 751 37 Uppsala, Sweden\label{aff159}
\and
Mathematical Institute, University of Leiden, Einsteinweg 55, 2333 CA Leiden, The Netherlands\label{aff160}
\and
Institute of Astronomy, University of Cambridge, Madingley Road, Cambridge CB3 0HA, UK\label{aff161}
\and
Univ. Lille, CNRS, Centrale Lille, UMR 9189 CRIStAL, 59000 Lille, France\label{aff162}
\and
Institute for Particle Physics and Astrophysics, Dept. of Physics, ETH Zurich, Wolfgang-Pauli-Strasse 27, 8093 Zurich, Switzerland\label{aff163}
\and
Department of Astrophysical Sciences, Peyton Hall, Princeton University, Princeton, NJ 08544, USA\label{aff164}
\and
Space physics and astronomy research unit, University of Oulu, Pentti Kaiteran katu 1, FI-90014 Oulu, Finland\label{aff165}
\and
Institut de Physique Th\'eorique, CEA, CNRS, Universit\'e Paris-Saclay 91191 Gif-sur-Yvette Cedex, France\label{aff166}
\and
International Centre for Theoretical Physics (ICTP), Strada Costiera 11, 34151 Trieste, Italy\label{aff167}
\and
Center for Computational Astrophysics, Flatiron Institute, 162 5th Avenue, 10010, New York, NY, USA\label{aff168}}    

\abstract{Higher-order correlation functions are firmly established as a fundamental tool for the statistical analysis of clustering in modern galaxy surveys. It was demonstrated that they greatly enrich the information content extracted by two-point statistics, allowing us to break the degeneracies between model parameters and constrain departures from Gaussianity.
This paper presents the statistical estimators adopted to evaluate the galaxy three-point correlation function and its numerical implementation within the data analysis pipeline of the Euclid Science Ground Segment. Two different algorithms are adopted to count  triplets: a direct and exact counting method capable of providing a robust three-point correlation function measurement for any triangular configuration, and a more efficient method based on spherical harmonic decomposition, designed to address the computational challenges of measuring the three-point statistics for data sets as large as those of the final \Euclid survey. The spherical harmonic decomposition estimates the Legendre coefficients of the three-point correlation function up to a finite expansion order. Despite being an approximation, the three-point function measured with this approach satisfies the scientific requirements of the mission. We also introduce, implement, and validate the random split technique, which reduces the computational cost of counting triplets in the reference random sample by a factor of 10, without significantly compromising numerical accuracy. We evaluated the robustness, precision, and accuracy of the numerical estimates through an extensive campaign of validation tests, the results of which are presented. Finally, we quantify the computational requirements and their scaling with the expected size of \Euclid data set, showing that a complete three-point analysis of the final \Euclid survey is within computational reach.}


\titlerunning{\Euclid preparation. Three-point correlation function estimation}
\authorrunning{Euclid Collaboration: Veropalumbo, A., et al.}

\maketitle
\nolinenumbers 

\section{\label{sc:Intro}Introduction}
Galaxy clustering (GC) investigates the large-scale structure (LSS) of the Universe as traced by galaxies. By exploring the relationship between galaxies and the underlying matter density, we can reconstruct the evolution of structures from earlier epochs to more recent times and gain insight into the Universe's foundational properties.
To efficiently extract information from the LSS, it is advantageous to measure
its summary statistics, representing the moments of the density field's likelihood.
The density field originates from stochastic initial conditions set by inflation and subsequently evolves under gravitational instability.
This methodology offers several benefits. First, the moments are marginalised across realisations,  thus detaching from the stochastic processes underlying the observed galaxy distribution.
Second, summary statistics allow straightforward comparisons with theoretical
predictions that hinge on cosmological parameters and galaxy properties. Lastly,
a clear hierarchy exists among the moments, with most information encapsulated in the first nonzero moment, two-point statistics. These are the two-point correlation function (2PCF), $\xi(r)$,
in configuration space, and its Fourier counterpart, the power spectrum, $P(k)$. Currently, most research focused on these two probes. In particular, experiments like the Baryon Oscillation Spectroscopic Survey \citep[BOSS,][]{Alam2017} and Dark Energy Spectroscopic Instrument \citep[DESI,][]{DESI1} measured two-point statistics of a huge collection of galaxies spanning a large volume. These observational efforts led to the use of the baryon acoustic oscillation (BAO) probe as a standard ruler \citep{Ross2017, Beutler2017a, DESI_VI}, as well as redshift-space distortions \citep[RSD; ][]{Satpathy2017, Beutler2017b} and direct cosmological interpretation \citep{Grieb2017, Sanchez2017}.

In the case of a Gaussian density field, the two-point statistics suffice to characterise the field fully.  However, the observed galaxy density field markedly deviates from a Gaussian distribution.
This non-Gaussian nature stems from several factors, predominantly the nonlinear
evolution of cosmic structures, which becomes more pronounced at later cosmic times. Other factors, including galaxy bias and RSD, contribute
significantly to this deviation. Additionally, primordial non-Gaussianities,
potentially originating immediately after the Universe's inflationary period,
could leave detectable signs in the galaxy density field.
These effects naturally lead to higher-order moments in the density field \citep[see e.g.][for a review]{Bernardeau2002}.

Higher-order statistics are essential for probing this regime and extracting complementary information on non-Gaussian physics, which may be challenging to assess with two-point statistics alone.
Moreover, higher orders can enhance cosmological constraints by breaking parameter degeneracies. Typical examples are the three-point correlation
function (3PCF), or its Fourier-space equivalent, the bispectrum.
So far, state-of-the-art works focused on higher-order analysis in Fourier space \citep{GilMarin2017, DAmico2020, DAmico2022a, Philcox2022, Cabass2022b, Cabass2022a, DESI_FSVII, DESI_FS, NovellMasot2025, Chudaykin2025} mainly because the bispectrum estimator is computationally more efficient and its modelling in the frequency domain is more tractable.

Exploiting the 3PCF presents several challenges.  Estimators for 3PCF are sourced by triplets \citep{Szapudi1998, Kayo2004}, thus facing computational complexities scaling as $\mathcal{O}(N^3)$.
This property makes it unfeasible to run 3PCF with standard approaches for densities and volumes covered by current and future surveys. Data partitioning algorithms, such as linked-list or $k$-d tree, partially mitigate this issue \citep[see e.g.][for discussion]{EP-delaTorre}, but challenges persist.
However, this approach was successfully applied to volume or density-limited samples such as SDSS \citep{Marin2011}, WiggleZ \citep{Marin2013}, and VIPERS \citep{Moresco2017} galaxy surveys, as well as to galaxy cluster catalogues \citep{Moresco2021}, achieving the first detection of the BAO peak in the 3PCF of galaxy clusters.

In the last ten years, to tackle the computational complexity of 3PCF, we witnessed the emergence of new techniques to extract this information more efficiently from large spectroscopic samples. \citet{Slepian2015} introduced a new triplet counting algorithm to compute the 3PCF in harmonic space. In particular, this approach guarantees algorithmic complexity $\mathcal{O}(N^2)$, thus significantly improving the direct triplet counting method. This comes together with the fact that, in general, the signal is well captured by a few multipoles, allowing for significant truncation without substantial information loss. Similar approaches can be generalised to measure the full anisotropic 3PCF \citep{Slepian2018, Sugiyama2019}. These advancements pushed the development of different independent tools to measure and exploit 3PCF data \citep{Marulli2016, Philcox2021, Sugiyama2023, Wang2023, Port2023, Farina2026, Labate2026}, demonstrating the current interest in the field.
This approach was applied to the BOSS survey \citep{Alam2017} for the analysis of the 3PCF over a wide range of triangular configurations, finally identifying the BAO feature in the galaxy 3PCF \citep{Slepian2017a, Slepian2017b, Kamalinejad2026}. The same technique was used for a joint 2PCF + 3PCF analysis \citep{Veropalumbo2021} to disentangle the linear growth rate and the clustering amplitude in the VIPERS survey \citep{Guzzo2014}. Similarly, this efficient technique simplified the development of theoretical studies aimed at quantifying the accuracy of our models for the 3PCF in real space \citep{Veropalumbo2022, Guidi2023} and in redshift space \citep{Kuruvilla2020, Sugiyama2021, Sugiyama2023, Farina2026, Pugno2024}.

\Euclid \citep{EuclidSkyOverview} is a space observatory tasked with constructing a comprehensive survey of the Universe. During its expected six-year lifespan, \Euclid will map a third of the sky at intermediate to large redshifts, charting the spatial distribution of galaxies across a 50~Gpc$^{3}$ volume. Thanks to the spectroscopic information, the survey will pinpoint the position of millions of galaxies in the cosmological volume, thus precisely determining the density field.
This vast data set will lead to high-quality galaxy clustering measurements. This will come in combination with the weak lensing probe, the focus of the photometric \Euclid survey. These observables will offer a unique and complementary view of the matter density field at different epochs. They will guarantee high-quality information to answer fundamental open questions in cosmology \citep{Blanchard-EP7}.

A key aspect of preparation for the mission involved developing highly accurate estimators of the clustering properties in Fourier and configuration spaces, capable of dealing with the large expected amount of data provided by the survey \citep[][Euclid Collaboration: Sefusatti et al. in prep.]{EP-delaTorre}. This paper details the implementation and validation of an efficient and accurate 3PCF estimator. This element was integrated into the Euclid Science Ground Segment (ESGS) pipeline, which oversees the full analysis from raw data to summary statistics. In addition, this algorithm supports the preliminary and future scientific 3PCF analyses, tailored for \Euclid-like scenarios \citep[][Euclid Collaboration: Pugno et al. in prep.]{EP-Guidi}.

This paper is organised as follows. After reviewing the theoretical background of the 3PCF estimators in Sects.\,\ref{sec:setup} and \ref{sec:estimator}, we describe the triplet counting algorithms in Sect.\,\ref{sec:triplets}. Section \ref{sec:pipeline} discusses the structure of the \Euclid pipeline.  In Sect.\,\ref{sec:validation} we discuss scientific validation and in Sect.\,\ref{sec:performances} we present computational performance, showing forecasts for the \Euclid mission. Finally, in Sect.~\ref{sec:conclusions} we draw our conclusions. In Appendix \ref{sec:appA} we report analytical formulae for bin-averaged Legendre polynomials. In Appendix \ref{sec:appB} we recall the formula for the theoretical covariance used in this analysis to calculate the reference error, and in Appendix \ref{sec:appC} we compare this work with other codes.

\begin{figure*}[t]
    \includegraphics[width=\textwidth]{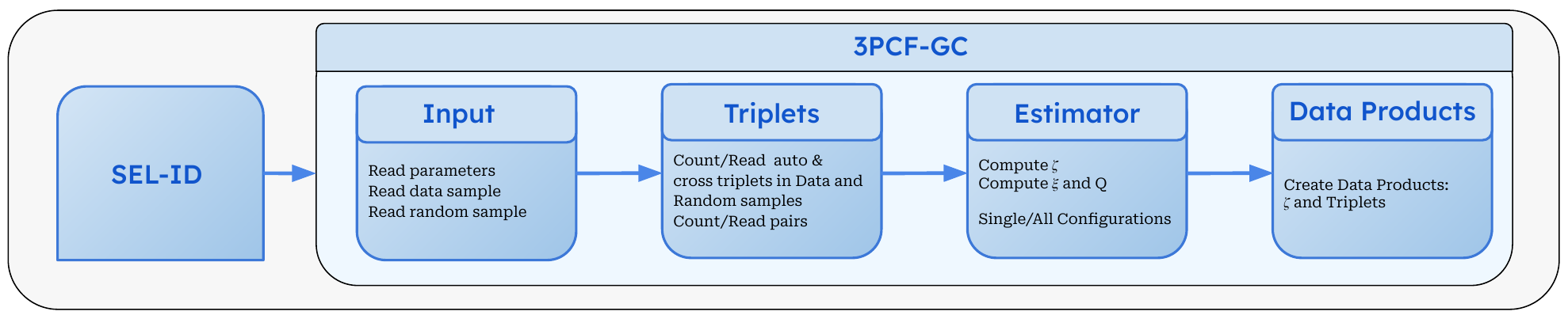}
    \caption{Schematic overview of the 3PCF-GC processing function and its integration within the ESGS pipeline. The upstream SEL-ID module provides the galaxy and random catalogues. The four processing steps are: input reading, triplet counting (auto- and cross-triplets for the data and random samples), 3PCF estimation (computing $\zc$, $\xi$, and the reduced 3PCF $Q$), and generation of the output data products in FITS format.}
    \label{fig:3pcf_pipeline}
\end{figure*}

\section{\label{sec:setup} Theoretical setup}
In principle, five independent variables are needed to characterise the 3PCF: three describing the triangle shape and two describing its orientation with respect to the line of sight. For \Euclid science cases, we only consider the isotropic 3PCF, where the two orientation degrees of freedom are averaged over uniformly.

The `connected' 3PCF, $\zc$, is an intrinsic statistical property of the density-contrast field, $\delta(\vec{x})$,
and it is defined as the ensemble average of the product of the density field in three different positions:
\begin{equation}
    \label{eq:zeta_def}
    \zc\VP = \ave{\delta(\vec{x_1})
         \delta(\vec{x_2})
          \delta(\vec{x_3})} \;,
\end{equation}
where $\langle\dots\rangle$ indicates the ensemble average and $\vec{r_{12}} = \vec{x_2} - \vec{x_1}$, $\vec{r_{13}} = \vec{x_3} - \vec{x_1}$.
The coordinates of the three vertices fully characterise the shape and orientation of each triangle.
Due to statistical homogeneity and isotropy, the number of independent variables
reduces to three: the sides of the triangle:
$\rab = \lVert \vec{x_1} - \vec{x_2} \rVert $, $\rac = \lVert \vec{x_1} - \vec{x_3} \rVert $,
and $\rbc = \lVert \vec{x_2} - \vec{x_3} \rVert $ or any related proxies.
As a consequence, the 3PCF can be parametrised in different ways exploiting the relation between triangle sides and subtended angles:
\begin{equation}
    \label{eq:side_triangle}
    \rbc^2 = \rab^2\,+\,\rac^2\,-\,2\,\rab\,\rac\,\ctheta,
\end{equation}
where $\theta$ is the angle between $\vec{r_{12}}$  and $\vec{r_{13}}$.
Our implementation offers the possibility of choosing between three different parametrisations:
\begin{itemize}
    \item {\tt SIDE}: $\{ \rab, \rac, \rbc \}$,
    \item {\tt THETA}: $\{ \rab, \rac, \theta  \}$,
    \item {\tt COSTHETA}: $\{ \rab, \rac, \ctheta  \}$.
\end{itemize}

In this section we use {\tt COSTHETA} parametrisation. Similar conclusions can be drawn when a different scheme is assumed.

As an alternative approach, it is also possible to consider 3PCF expansion in harmonic space
\begin{equation}
    \label{eq:zeta_ell_def}
    \zc\CP = \sum_{\ell=0}^{\infty} \zlF \legG(\ctheta) \;,
\end{equation}
where $\zlF$ are the Legendre coefficients of the 3PCF, expanded for the angle of the triangle aperture,
and $\legG$ is the Legendre polynomial of degree $\ell$.
\citet{Slepian2015} first presented a technique to measure this quantity.
In the next sections we evaluate the impact of truncating the series and estimate the maximum multipole $\lmax$.

We introduce the `binned' 3PCF, which is the average of  $\zc$ over bins of size $\Delta \rab$, $\Delta \rac$,
\begin{align}
    \label{eq:zeta_binned}
    \bzc\CP = \int_{\vec{V_{12}}} \int_{\vec{V_{13}}} \diff^3 \vec{r}_1  \, \diff^3 \vec{r}_2 \, \zc \VP \Theta(\vec{r_1}, \vec{r_2}; \cos \theta) \;,
\end{align}
where $V_{12}$ and $V_{13}$ denote the spherical shells of radii $\rab \pm \Delta \rab /2$ and $\rac \pm \Delta \rac /2$, respectively, and $\Theta(\vec{r_1}, \vec{r_2}; \cos \theta)$ is the binning function, with different parametrizations to characterize triangles.
For the {\tt COSTHETA} and {\tt THETA} parametrisations, $\Theta$ reduces to a normalised indicator function on the angular variable, $\Theta = \delta_{\rm D}(\ctheta - \mu)/\Delta\mu$ or $\Theta = \delta_{\rm D}(\theta - \theta_0)/\Delta\theta$, respectively. For the {\tt SIDE} parametrisation, $\Theta$ selects configurations whose third side $\rbc$ falls within the specified bin; the resulting integration is more involved and is discussed in Appendix~\ref{sec:appA}.

By plugging Eq.~\eqref{eq:zeta_ell_def} into Eq.~\eqref{eq:zeta_binned} and integrating,
we determine the link between the binned 3PCF and the binned coefficients of the expansion
\begin{align}
     \label{eq:zeta_ell_binned_def}
    \bzc\CP = \sum_{\ell=0}^{\infty} \bar{\zeta}_{\ell}(r_{12}, r_{13}) \bar{\mathcal{L}}_{\ell}(r_{12}, r_{13}, \theta) \;,
\end{align}
where $\blegG$ are the triangle-binned Legendre multipoles.
The expression for $\blegG$ is trivial for parametrisations of types {\tt THETA} and {\tt COSTHETA}.
The {\tt SIDE} binning is more complicated and involves a double integration (see Appendix \ref{sec:appA}).

Another useful quantity is the `reduced' 3PCF
\begin{align}
    \label{eq:red_3pcf}
    Q\CP \, = \, \frac{\zc\CP}{\xi(\rab)\xi(\rac) + \xi(\rab)\xi(\rbc) + \xi(\rac)\xi(\rbc)} \; ,
\end{align}
where $\xi$ is the isotropic 2PCF computed on the sides of the triangle, and $\rbc$ is calculated from Eq.~\eqref{eq:side_triangle}.
This quantity is useful in some circumstances, for example
when we want to neglect the absolute 3PCF normalisation, and for this reason is also estimated in our 3PCF implementation.

\section{\label{sec:estimator} The 3PCF estimator}
The probability of finding a triplet of galaxies at the vertices of a triangle of sides $\rab, \, \rac, \, \rbc$
in a population is given by
\begin{align}
\label{eq:zetac_def}
\diff P_{123} = \bar{n}^3 \,  \diff V_1 \, \diff V_2 \, \diff V_3 & \left[1 + \xi(\rab) + \xi(\rac) + \xi(\rbc) \right. \nonumber \\
&\quad\left. +  \zc\SP \, \right] \;,
\end{align}
where $\zc$ denotes the connected 3PCF, $\xi$ is the 2PCF, $\bar{n}$ is the mean galaxy density, and $\diff V_{i}$ are the volume elements. The connected 3PCF measures the intrinsic probability of three galaxies forming a triangle, beyond what is expected from the mean density and from two-point correlations alone. To isolate this contribution from triplet counts, one must subtract the disconnected part: the triplets formed by random objects, or by a random object together with a correlated pair.
We use the estimator for the connected 3PCF proposed by \citet{Szapudi1998}
\begin{equation}
\label{eq:SS98_expanded}
    \ezc\SP \, = \, \frac{\mathrm{DDD} - 3\,\mathrm{DDR} + 3\,\mathrm{DRR} - \mathrm{RRR}}{\mathrm{RRR}} \;.
\end{equation}
where  the quantities $\mathrm{DDD}$, $\mathrm{DDR}$, $\mathrm{DRR}$, and $\mathrm{RRR}$ denote the counts of triplets in configuration space, normalised to the total number of triplets.

The data sample $D$ contains the signal, while the random catalogue $R$ serves as a Monte Carlo sampling of the survey volume, accounting for geometric
and selection effects. The denominator $\mathrm{RRR}$ measures the random probability of finding a triplet within a given volume. This aspect of the analysis is crucial to accurately determine the clustering statistics.
The generation of the random sample for the \Euclid survey data will be carefully estimated by a specific part of the ESGS pipeline (Visibility Mask SPectroscopy, VMSP) and is described in detail in the paper Euclid Collaboration: Granett et al. (in prep.).
The measured 3PCF is a stochastic variable. \citet{Szapudi1998} showed that Eq.~\eqref{eq:SS98_expanded} is unbiased and minimum-variance: its expectation value recovers the true 3PCF, and the variance from sample discreteness is minimised by removing edge effects. When deriving the variance expressions, \citet{Szapudi1998} assumed a continuous limit for survey-volume sampling. This may not always be accurate: to reduce discreteness effects, the random density is chosen to be significantly higher than that of galaxies. The random sample generated by the VMSP processing element is \num{50} times denser than the galaxy catalogue to ensure sub-percent precision in estimating the two-point statistics. This choice is driven by the precision requirements of the 2PCF estimator. As no specific precision requirement was imposed for the 3PCF, the same random catalogues were adopted for all clustering measurements in this study. However, such a high density of random objects would render the estimation of the 3PCF for a \Euclid-like sample computationally infeasible. To address this challenge, we implemented the random-split technique described in Sect.~\ref{sec:random_split}.

\section{\label{sec:triplets} Algorithm description}

This section describes the options available for estimating the 3PCF. We first introduce the triplet-counting methods that form the backbone of the estimate, then the triangle-configuration schemes for which the 3PCF is evaluated, and finally the random-splitting technique, detailed in \citet{Keihanen2019}.

A direct evaluation of Eq.~\eqref{eq:SS98_expanded} would require counting auto- and cross-triplets ($\mathrm{DDD}$, $\mathrm{DDR}$, $\mathrm{DRR}$, $\mathrm{RRR}$) separately and combining them. We instead follow the procedure of \citet{Slepian2015}, which works on a single combined catalogue. We define
\begin{equation}
    \label{eq:joint_field}
    N = D \cup R_{\alpha} \,,
\end{equation}
where $R_\alpha$ denotes the random catalogue with each object reweighted by $-\alpha$. The factor $\alpha$ is fixed by requiring the total weight of $N$ to vanish:
\begin{equation}
    \label{eq:random_weight}
    \alpha = \frac{\sum_{i=1}^{N_{\rm D}} w^{\rm D}_{i}}{\sum_{i=1}^{N_{\rm R}} w^{\rm R}_i} \,,
\end{equation}
with $N_{\rm D}, N_{\rm R}$ the number of objects in $D$ and $R$, and $w_i^{\rm D}, w_i^{\rm R}$ their weights. Counting weighted triplets in $N$ directly produces the numerator of Eq.~\eqref{eq:SS98_expanded}, so the estimator reduces to
\begin{equation}
    \label{eq:SS98_compact}
    \ezc\SP \, = \, \frac{\mathrm{NNN}}{\mathrm{RRR}} \,.
\end{equation}
This formulation simplifies bookkeeping by eliminating the need to track the four triplet types independently. For the SHD method (Sect.~\ref{sec:SHD}), it is in fact the only viable approach because the harmonic coefficients are computed from the combined catalogue by construction.

\subsection{\label{sec:direct_counts} Direct triplet counting}
The first method we consider is direct counting (hereafter DC).
The first step of this brute-force method is to search for all neighbours around a primary particle up to a specific separation. The neighbours are split into different radial bins around the primary, and then these bins are cross-correlated two-by-two, forming all possible triangles of side $\paren{\rab, \rac, \rbc}$.
The procedure is then iterated for each object in the sample, considered as the primary vertex of the triangle.
We count triplets in both the $N$ and $R$ samples. We then apply Eq.~\eqref{eq:SS98_compact} to estimate $\zc$.

This simple algorithm guarantees that all triplets, including objects of all types, are included in the counts. Its downside is the computational cost, which scales as $O(N^3)$ where $N$ is the number of objects in the samples. Although data partition schemes such as $k$-d tree and linked lists \citep[see also][for a discussion]{EP-delaTorre} can be implemented to reduce computational burden, the overall computational cost remains prohibitive for a data set as large as the \Euclid survey. Yet, we chose to include the DC option in the numerical code, since it provides a useful benchmark to assess the quality of the results obtained with the other algorithm and since it could also be used to evaluate the 3PCF of some specific types of rare objects, like QSOs and galaxy clusters.

\subsection{\label{sec:SHD} Spherical Harmonic Decomposition}
The second method, called Spherical Harmonic Decomposition (hereafter SHD), was introduced in \citet{Slepian2015}, to which we direct readers who are interested in the numerical and mathematical details. It substantially improves efficiency in triplet counting.
As in the DC case, the algorithm iterates over all primary particles in the catalogue, identifying and binning the neighbour counts in radial shells of fixed thickness and increasing radii at different separations up to a predefined maximum distance. Then, the local density field around each primary is expanded on the basis of spherical harmonics $Y_{\ell m}$, obtaining a set of coefficients $a_{\ell m}$ for each shell. For a discrete catalogue, the coefficients in the shell at distance $\rab$ from a primary at position $\vec{s}$ are given by
\begin{equation}
    \label{eq:alm_discrete}
    a_{\ell m}(\rab;\,\vec{s}) \;=\; \sum_{j \,\in\, V_{12}(\vec{s})} w_j \; Y_{\ell m}^*(\hat{\vec{r}}_{sj}) \;,
\end{equation}
where the sum runs over all objects $j$ within the shell $V_{12}(\vec{s})$ centred on $\vec{s}$, $w_j$ is the object weight, and $\hat{\vec{r}}_{sj}$ is the unit vector from the primary to object $j$ \citep[see][for the full derivation]{Slepian2015}. The information is then compressed to a chosen angular resolution set by the maximum multipole index $\lmax$.
These coefficients are then cross-correlated to obtain the Legendre coefficients $T_{\ell}$ of the triplets around the position $\vec{s}$ of the primary particle
\begin{equation}
    \label{eq:triplets_around_primary}
    T_{\ell}(\rab, \rac; \vec{s}) = \frac{1}{4\pi} \delta(\vec{s}) \sum_{-m}^{m} a_{\ell m}(\rab;\vec{s}) \, a_{\ell m}^* (\rac;\vec{s})\;,
\end{equation}
where $T_{\ell}$ are the Legendre coefficients of the triplets around a primary at position $\vec{s}$, $\delta(\vec{s})$ is the density contrast field at $\vec{s}$ and $a_{\ell m}$ are the spherical harmonics coefficients of the density field in the shell at distance $r$ from the primary particle in $\vec{s}$. In general, two sides and three indices are needed to characterise the triangle and its orientation with the line of sight \citep{Slepian2018, Sugiyama2019}. Statistical isotropy, which we assume here, allows us to compress the information into a single index. We then average Eq.~\eqref{eq:triplets_around_primary} over the survey volume (that is, we loop over all primary particles) to get the total Legendre coefficients of the triplet counts.
Since the computation of $a_{\ell m}$ is performed for each shell separately, the algorithm complexity scales to $\mathcal{O}(N^2)$, a substantial improvement from the previous case.
It should be stressed that the accuracy of the estimate depends on the maximum multiple moment considered in the expansion $\lmax$. We will test the sensitivity of the results to $\lmax$ in Sect.~\ref{sec:shd_conv}. The procedure is repeated for mixed and random samples, obtaining Legendre multipoles of triplet counts for the $N$ sample $\vec{\mathcal{N}} = \curly{\mathcal{N}_0, \dots, \mathcal{N}_{\ell_{max}}}$ and for the $R$ sample $\vec{\mathcal{R}} = \curly{\mathcal{R}_0, \dots, \mathcal{R}_{\ell_{max}}}$.

The full 3PCF can be estimated from the multipoles in two ways.
The first method, which we name direct sum $\hat{\zeta}_{\rm DS}$, consists of independently reconstructing the triplet counts in triangle space using Eq.~\eqref{eq:zeta_ell_def} and then applying the estimator in Eq.~\eqref{eq:SS98_compact}
\begin{equation}
    \label{eq:zeta_ds}
    \hat{\zeta}_{\rm DS}\CP = \frac{\sum_{\ell=0}^{\lmax} \mathcal{N}_{\ell} \paren{\rab, \rac} \legG(\ctheta) } {\sum_{\ell=0}^{\lmax} \mathcal{R}_{\ell} \paren{\rab, \rac}\legG(\ctheta) }\;,
\end{equation}
where we write for simplicity the 3PCF in {\tt COSTHETA} parametrisation.

As an alternative, we consider using the modified Szapudi--Szalay estimator to obtain the 3PCF's Legendre coefficients directly \citep[see][for detailed computation]{Slepian2015}.
This estimator for 3PCF Legendre multipoles, which we name here $\vec{\hat{\zeta}}_{\rm SE}$, reads
\begin{equation}
    \label{eq:zeta_se}
    \vec{\hat{\zeta}}_{\rm SE}\paren{\rab, \rac} = \paren{\tens{I} + \tens{M}}\inv  \frac{\vec{\mathcal{N}}\paren{\rab, \rac}}{\mathcal{R}_0\paren{\rab, \rac}} \;,
\end{equation}
where $\tens{I}$ is the identity matrix,  $\vec{\hat{\zeta}}_{\rm SE} = \curly{\mathcal{\hat{\zeta}}_0, \dots, \mathcal{\hat{\zeta}}_{\ell_{\rm max}}}$ is the vector of expansion coefficients for the 3PCF while $\vec{\mathcal{N}}$ is the corresponding vector of coefficients for the object counts for sample $N$. The correction for the edge effects comes in two parts: an isotropic normalization $\mathcal{R}_0$, which is the monopole ($\ell=0$) of the random triplet counts and hence a scalar,
and a mode-mixing matrix $\tens{M}$.  The latter quantity is computed from the Legendre multipoles of the triplet counts from the $R$ sample $f_{\ell} = \mathcal{R}_{\ell}/\mathcal{R}_0$ as
\begin{equation}
    M_{\ell,k} \, = \, \paren{2k+1}  \sum_{\ell^{\prime}>0}
    \begin{pmatrix}
\ell & \ell^{\prime} & k \\
0 & 0 & 0
\end{pmatrix}^2 f_{\ell^{\prime}} \;,
\label{eq:mixing_matrix}
\end{equation}
where the term in parentheses is the Wigner 3$j$ symbol.
From the definition of $f_{\ell}$, it is evident that $\tens{M}$ is scale dependent; we dropped the dependence $\paren{\rab, \rac}$ from Eq.~\eqref{eq:mixing_matrix} for clarity. This term is zero in a uniform survey with no boundaries or selection effects.
The two methods coincide for $\lmax\rightarrow\infty$. Equation~\eqref{eq:zeta_ds} reconstructs the full 3PCF in triangle space from the harmonic coefficients and then applies the Szapudi--Szalay estimator, while Eq.~\eqref{eq:zeta_se} directly yields the 3PCF Legendre multipoles by absorbing the edge correction into an isotropic normalisation $\mathcal{R}_0$ and a mode-mixing matrix $\tens{M}$. The two quantities are derived from the same estimator; at finite $\lmax$, however, they may differ because the truncation of $\tens{M}$ is not required to match that of the data. In practice, the agreement between the two methods should be verified, especially in the presence of complex survey windows or for small fields, where edge effects are more pronounced. In such cases, a comparison with the DC estimator, which is exact by construction, provides a valuable cross-check when computationally feasible. We consider both strategies in our implementation and compare them in Sect.~\ref{sec:edgecorr}.

\subsection{\label{sec:single_all} 3PCF triangle configurations}
For both 3PCF estimators described previously, we offer two options to output results.
In the first option, named `all configurations', we count all possible triplets in all triangle configurations from minimum to maximum separation in a given binning $\Delta r$. For this option, we use the SIDE parameterisation defined in Sect.~\ref{sec:setup}.
The output consists of a set of $\hat{\zc}(\rab, \rac, \rbc)$ defined for every possible combination of triangular sides.
In harmonic space, the corresponding result is the whole set of coefficients $\zlF$ for every combination of $\rab$ and $\rac$ and up to a given $\lmax$.

The second option, called `single configuration', selects a subset of triangle configurations by fixing the first two sides of the triangle and the binning, which are inputs to the code. Unlike in the previous case, the user can select any of the three possible parametrisations. This second option offers the possibility of selecting the size of the bin and the triangle configuration, which ultimately allows one to focus on specific triangle configurations and to test the sensitivity of the results to the choice of the bin and of $\lmax$.

\subsection{\label{sec:split_description} Random split}
To calculate the 3PCF, we have to combine two samples: the data catalogue, which contains the signal, and the random catalogue, used to correct for selection effects. Both samples are affected by a shot noise due to the finite number of objects in both catalogues.
While the shot noise of the data sample is given, the shot noise of the random sample can be minimised by increasing the number of random objects, $N_{\rm R}$. For the \Euclid survey, $N_{\rm R}$ is required to be at least \num{50} times greater than the number of objects in the data sample $N_{\rm D}$.
The triplet count algorithm dominates the computational efficiency of the 3PCF measure. This becomes even more critical when computing random triplets. For this reason, we implement the random split techniques, first introduced for the 2PCF only in \citet{Keihanen2019}. This technique consists of splitting the random sample into many realisations $R_i$, of smaller density, and computing the triplets in the samples $R_i$ and in the joint catalogues $N_i = D \cup R_{\alpha,i}$ as defined in Sect.~\ref{sec:triplets}. All these measurements are then averaged to estimate 3PCF using Eq.~\eqref{eq:SS98_compact}

\begin{equation}
    \label{eq:zeta_split}
    \hat{\zeta}_{\rm split} \CP \,  = \, \frac{\sum_{i=1}^{N_{\rm split}} \mathrm{NNN}_i}{\sum_{i=1}^{N_{\rm split}} \mathrm{RRR}_i} \;,
\end{equation}
where $\hat{\zeta}_{\rm split}$ is the 3PCF estimated with the split method and $N_{\rm split}$ is the number of subsamples in which we choose to split the random catalogue.

This technique allows for the gain of up to a factor of \num{10} in the 3PCF computation using the SHD method, which becomes much larger for DC. We verify that for 3PCF the optimal choice for the split fraction is between $ 1.5 < N_{\rm R_i}/N_{\rm D} < 2$, in agreement with \citet{Slepian2015} and \citet{Keihanen2019}. This value was extrapolated following the argument presented in \citet{Slepian2015}; we refer the reader in particular to their Sect.~5 for an extended discussion. For a random sample \num{50} times larger than the data sample, this corresponds to $N_{\rm split} = 25$. We validate this methodology in Sect.~\ref{sec:split}, and assess the computational performance in Sect.~\ref{sec:random_split}.
For a similar discussion of the 2PCF, we refer the reader to \citet{EP-delaTorre}.
Throughout this work, where not stated explicitly, we use this method as our baseline for the computation of the 3PCF.

\section{\label{sec:pipeline} 3PCF and the \Euclid spectroscopic pipeline}
The 3PCF-GC Processing Function (PF) is a numerical code written in C++, developed in alignment with the ESGS standards to maintain consistency in code development, data storage, and computation, regardless of the machine that runs the code \citep{Frailis2019}. This development adhered to the Euclid Development Environment (EDEN), which dictates that libraries and their respective versions be used by all software developed within the ESGS, thus preventing inconsistencies or functionality changes in development and production phases. 3PCF-GC was integrated into the COllaborative DEvelopment ENvironment (CODEEN), a continuous integration and delivery (CI/CD) platform that automates the building, unit testing, and distribution of all scientific software in the SGS. The source code is managed via a version control system (GitLab) and can be executed through a CI/CD pipeline, finally being deployed on a distributed file system accessible at all SDCs. This infrastructure ensures smooth and effective SGS operations across all SDCs but is also capable of stand-alone execution, which is essential for testing and validation processes.

The 3PCF-GC processing follows four main steps, as shown in Fig.~\ref{fig:3pcf_pipeline}. Initially, it reads inputs that include a configuration file, data, and random catalogues, and optionally pre-computed triplet counts if specified in the configuration file, if available. In pipeline mode, these catalogues are supplied by the preceding SEL-ID processing function in the SGS pipeline. SEL-ID extracts a catalogue from the Euclid Wide Survey (EWS) based on specific selection criteria and provides it, along with the corresponding random catalogue, to 3PCF-GC. The input galaxy and random catalogues are then read and organised into a linked list.

Next, internal data structures are built, and the counting algorithm scans the primary galaxies to identify triplets according to the chosen configuration. The weighted triplet counts for each triangular configuration bin are stored in arrays. Depending on the 3PCF estimator selected by the user, 3PCF-GC performs the necessary triplet counts in series or reads them from the input files. These triplet counts are then combined to compute the 3PCF estimate. Finally, the 3PCF and individual triplet count products are prepared and delivered as FITS files \citep{Pence2010}.

\section{\label{sec:validation} Validation tests}
In this section, we present the validation of the \Euclid 3PCF estimator.
We compare the performance of the different 3PCF options and discuss the best choice for application to a \Euclid-like data set.

\begin{figure*}[t]
    \centering
    \includegraphics[width=\linewidth]{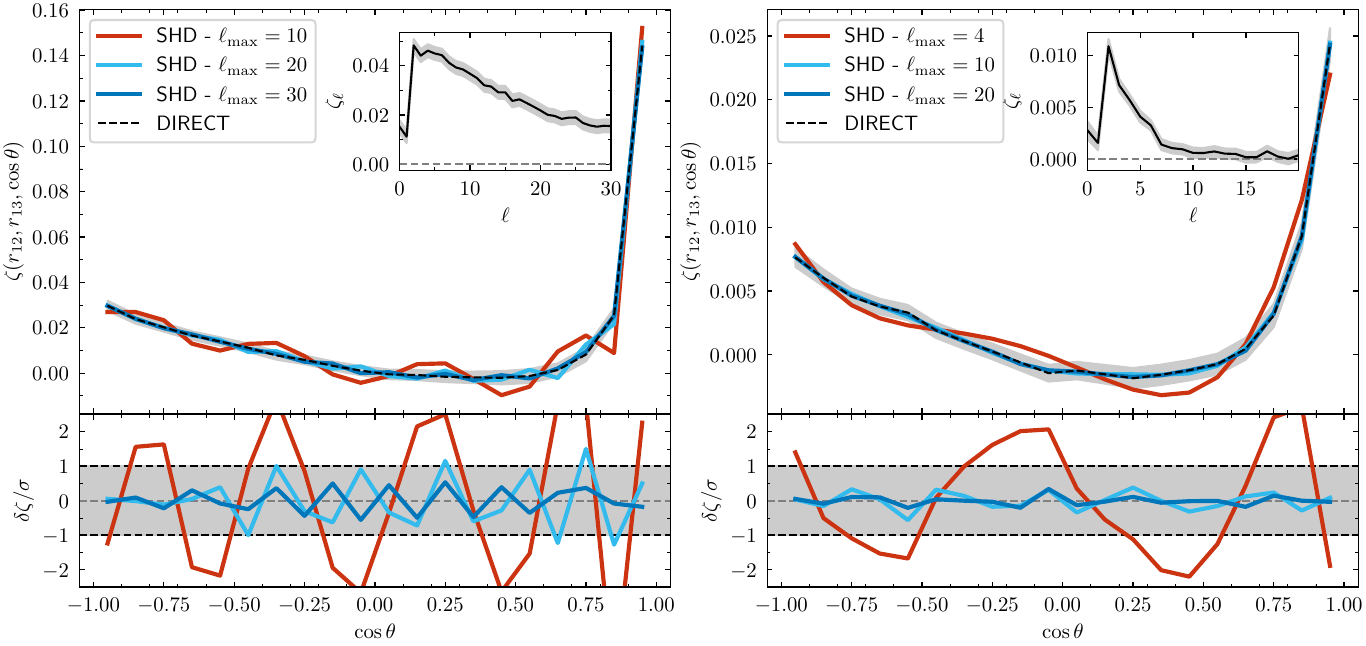}
    \caption{Convergence of the SHD method as a function of $\cos\theta$ for two triangle configurations from the Flagship mock in the first redshift bin. \emph{Left:} Isosceles case ($\rab = \rac = 20\,\protect\hMpc$) with $\lmax = 10$ (red), $20$ (cyan), and $30$ (blue), compared to the DC reference (black dashed). \emph{Right:} Non-isosceles case ($\rab = 20\,\protect\hMpc$, $\rac = 40\,\protect\hMpc$) with $\lmax = 4$ (red), $10$ (cyan), and $20$ (blue). The grey bands show the expected DR1 statistical error from the analytical covariance. Lower panels display the difference between the SHD and DC estimates in units of the statistical error. Insets show the Legendre coefficients $\zeta_\ell$ as a function of $\ell$, illustrating the slow convergence for the isosceles case and the rapid convergence for the non-isosceles one.}
    \label{fig:iso_noniso}
\end{figure*}

\subsection{Data sets}
\label{sec:data}

To perform the validation tests presented in this section, we used two different data sets: a realistic one, obtained from the Flagship Galaxy Mock \citep{EuclidSkyFlagship}, and an ideal catalogue of objects with zero three-point correlation signal.

\begin{description}
    \item[{\bf Flagship Galaxy Mock}:]
    the Flagship Galaxy Mock is a large catalogue of galaxies distributed in an octant of the sky \citep[for details see][]{EuclidSkyFlagship}. This huge collection of data serves different purposes, including the one for spectroscopic galaxy clustering analyses. In fact, we extracted a sample of galaxies with flux \ha~greater than $2\times 10^{-16}\,\mathrm{erg}\,\mathrm{s}^{-1}\,\mathrm{cm}^{-2}$ within a \SI{2500}{\deg\squared} area and in the redshift range $0.9 \leq z\leq 1.1$. The chosen area is $1/6$ of the expected total \Euclid survey and roughly corresponds to the area of the Euclid Data Release 1 (EDR1). We also apply the expected completeness cut of 0.43 \citep{Blanchard-EP7}. The final sample has $6\times 10^5$ objects. We obtained similar selections for the redshift bins $1.1<z<1.3$, $1.3<z<1.5$, $1.5<z<1.8$, mimicking the observation strategy of the \Euclid spectroscopic survey. Although we performed the tests using the samples in all redshift bins, we will show only the results obtained in the redshift range $[0.9,1.1]$. This choice is motivated by the similarity of the results and by the fact that the number density of galaxies in the bin, which is the largest in the survey, allows us to stress-test the computational requirements of our code.
    \item[{\bf Gaussian mocks}:]
    The Gaussian mocks consist of a synthetically generated data set. These simulations have a non-zero two-point correlation signal and no higher-order correlation signals. We use \textsc{COVMOS} \citep{Baratta2023}, a \textsc{Python} package to create fast simulations with a given cosmology. The density field in Fourier space is generated on a grid using a Gaussian process with zero mean and variance given by the desired power spectrum. We ensure that the variance is small enough to respect the constraint that $\delta\geq-1$. Through a Fourier transform, we obtain the $\delta$ field in configuration space. Given the desired mean density, the particles are sampled point-by-point from this field using a Poissonian process. We generated \num{600} Gaussian mocks in a box with number density and within a volume similar to the Flagship Galaxy Mock. The result of this process is sensitive to the chosen parameters, particularly the number of cells per side used in generating the density field. To account for this effect, we generated three data sets with different numbers of cells ($N_{\rm cell} = 128^3,\,256^3,\,512^3$). Except for the clustering amplitude, $\sigma_8$, all catalogues were generated assuming the same cosmological model as the one adopted for the Flagship simulation.
\end{description}

\subsection{Modelling 3PCF covariance matrix}
\label{sec:covariance}

To model the 3PCF uncertainties and their covariance, we adopt an analytic model based on the assumption that (i) the galaxy density field is Gaussian and (ii) that the effect of the survey geometry, and consequently its window function, can be ignored \citep{Slepian2015}. With these assumptions, the model for the 3PCF covariance matrix solely depends on the survey volume, number density of the sources, and the galaxy 2PCF. The combination of these three quantities and measurement specifications, such as binning and triangle sides, fully characterises the theoretical prediction.

Gaussian errors are likely to underestimate those of a real survey in which departures from Gaussianity, boundary, and selection effects cannot be neglected \citep{Veropalumbo2022}. To minimise the impact of these effects we restrict our analysis on a range of scales from \SIrange[range-phrase={~to~}]{30}{140}{\hMpc}. This choice excludes small separations, in which the Gaussian assumption breaks down, and large separations, which are more affected by edge effects. In Appendix \ref{sec:appB}, we provide a detailed account of the model used in the analysis and validation tests. These tests are compared with a numerical covariance derived from $N$-body simulations.
The numerical code to generate the 3PCF Gaussian error model used in this paper is publicly available within the \textsc{MElCorr}\footnote{\url{https://gitlab.com/veropalumbo.alfonso/melcorr}} library to model clustering statistics and their covariance errors.

\subsection{\label{sec:shd_conv}Method comparison}

\begin{figure}[t]
    \centering
    \includegraphics[width=\linewidth]{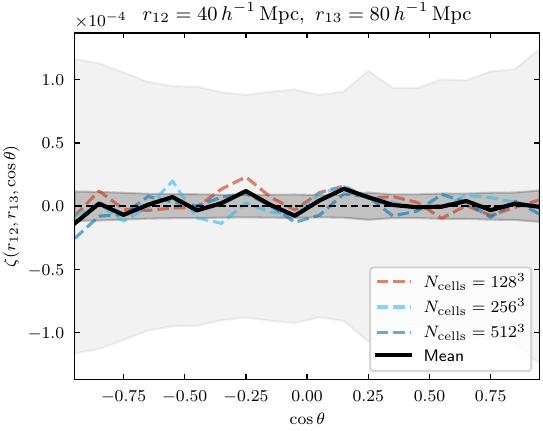}
    \caption{Null test: the mean 3PCF measured from \num{600} Gaussian mocks with the expected volume and density of DR1, for a scalene configuration with $\rab = 40\,\protect\hMpc$ and $\rac = 80\,\protect\hMpc$. The expected signal is zero by construction. The light-grey band represents the single-realisation statistical error; the dark-grey band shows its \SI{10}{\percent} amplitude, corresponding to the \Euclid systematic requirement. The solid black line is the mean over all mocks. Dashed lines show the mean for sub-samples generated with different grid resolutions, as indicated in the legend.}
    \label{fig:gaussian}
\end{figure}

\begin{figure*}[t]
    \centering
    \includegraphics[width=\linewidth]{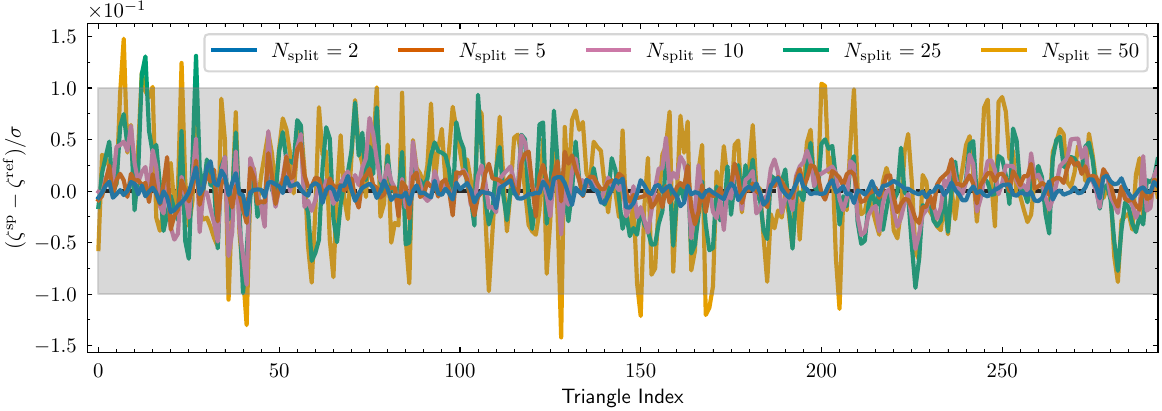}
    \caption{Impact of the random split on the 3PCF accuracy. We show the difference between the split and reference (no-split) estimates, normalised to the expected statistical error, for all non-isosceles triangle configurations with sides in the range $20$--$150\,\protect\hMpc$. The measurement uses the SHD method applied to the Flagship~2 catalogue reproducing the DR1 of \Euclid in the first redshift bin, with a random sample \num{50} times denser than the data. Each triangle index identifies a unique ($\rab$, $\rac$, $\rbc$) configuration with $\rab > \rac \geq \rbc$. Different colours correspond to different split factors $N_{\rm split}$, as indicated in the legend. The grey band marks the $\pm 10\%$ systematic requirement.}
    \label{fig:split-DR1}
\end{figure*}

\begin{figure}[t]
    \centering
    \includegraphics[width=\linewidth]{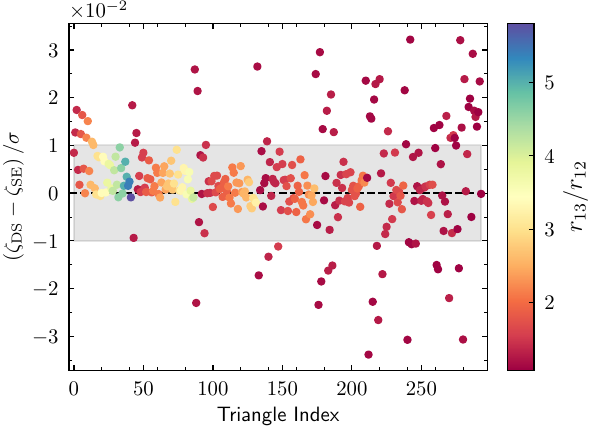}
    \caption{Comparison of the two edge-correction methods (direct sum $\hat{\zeta}_{\rm DS}$ and Slepian--Eisenstein estimator $\hat{\zeta}_{\rm SE}$) for the Flagship mock. The difference is shown in units of the expected statistical error as a function of the triangle index. Points are colour-coded by the ratio $\rac/\rab$, ranging from nearly isosceles configurations (warm colours) to more elongated ones (cool colours). The grey band marks the $\pm 1\sigma$ region. The triangle scale increases from left to right.}
    \label{fig:edgecorr}
\end{figure}

Both triplet counting methods considered here, DC and SHD, are exact; the first works directly in triangle space, while the second evaluates harmonic coefficients. However, in practice, for the SHD case, only a finite number of coefficients can be estimated, up to a maximum multipole $\lmax$. As a result, the infinite series in Eq.~\eqref{eq:zeta_ell_def} is truncated at $\lmax$, leading to an approximate estimate of the 3PCF. The precision of this estimate depends on how rapidly these coefficients $\zeta_{\ell}$ approach zero \citep{Slepian2015, Veropalumbo2021}, which, in turn, depends on the specific triangle configuration. This outcome does not depend on the statistical properties of the sample but only on the type of signal we aim to capture through the expansion in Legendre polynomials. For this reason, the convergence of the SHD method should be tested on a case-by-case basis. In the following, we present two illustrative cases: one in which the expansion achieves convergence and another in which it does not. In the left panel of Fig.~\ref{fig:iso_noniso}, we show the 3PCF of the Flagship catalogue in the first redshift bin for an isosceles configuration case ($r_{12} = r_{13} = 20 \, \hMpc$).  In the plot, we compare the result obtained with the DC method (black curve) with those obtained with the SHD method using three different values of $\lmax=10, 20, 30$ (red, cyan, and blue curves, respectively). The grey band represents the theoretical standard deviation expected from DR1 in the \Euclid catalogue, described in Sect.~\ref{sec:covariance}. In the bottom panel, we show the differences between the 3PCF estimated with the SHD method and that obtained from the DC method, taken as a reference, for each of the three $\lmax$ values considered. The accuracy increases with $\lmax$, as expected. However, even with the largest $\lmax$ value considered (cyan), the discrepancies in the reference case remain comparable in magnitude to the expected statistical error. This happens because truncating the Legendre expansion does not capture the steep rise that occurs at $\cos(\theta) \rightarrow 1$. Indeed 3PCF in harmonic space retains significant signals even for large values of $\ell$, and an expansion up to very high $\ell$ is needed to reach full convergence, as highlighted by the inset of the figure.
This behaviour is common to all configurations in which the third side $\rbc$ approaches $0$. At small scales, non-linearities give rise to a large 3PCF signal that then rapidly decreases towards larger separations. The isosceles configuration is a perfect example, and the problem is more severe the larger the value of $r_{12}$.

We repeated the same test for a scalene triangle configuration case ($r_{12} = 20 \hMpc$, $r_{13} = 40 \hMpc$). The result is presented in the right panel of Fig.~\ref{fig:iso_noniso}, analogously to the one on the left. The accuracy of the SHD estimate is now significantly higher, even for moderate values of $\lmax$. The reason for this improvement is that the expansion coefficients $\zeta_{\ell}$ rapidly approach zero, as shown in the inset, which leads to a much better agreement with the reference DC case.

These examples show that the most challenging cases occur when $r_{12} \approx r_{13}$. The reason is that, for these cases, given the triangle closure relation shown in Eq.~\eqref{eq:side_triangle}, the 3PCF is calculated for scales ranging from $r_{23} = 0 \, (\cos{\theta}=1)$ to $r_{23} = 2r_{12} \, (\cos{\theta}=-1)$. This implies that the 3PCF is typically $\gg 1$ when approaching $r_{23}=0$ and then rapidly drops to zero. Therefore, this rapid variation, which becomes progressively more severe for triangles of increasing size, is challenging to capture with the multipole expansion. In contrast, no large variations are expected when $\rac \gg \rbc$, resulting in an accurate estimate of the 3PCF for moderate values of $\lmax$.

These results indicate that, for certain triangle configurations, the 3PCF estimated using the SHD method may not meet the stringent accuracy requirements typically adopted for two-point statistics, namely that systematic uncertainties should be less than $10\%$ of the expected statistical error. However, this does not necessarily have a negative impact on the evaluation of the cosmological parameters since the likelihood analysis can be performed by comparing the measurement and the model of the individual multipoles rather than the full 3PCF \citep[see e.g.][]{Slepian2015, Slepian2018, Veropalumbo2021, Veropalumbo2022, Guidi2023, Pugno2024}. In fact, this approach is also preferred from a theoretical point of view, since individual multipoles are considerably easier to model than the full 3PCF \citep[][Euclid Collaboration: Pugno et al. in prep.]{EP-Guidi}.
From a practical standpoint, the signal-to-noise ratio of the 3PCF saturates rapidly with $\lmax$ for non-isosceles configurations, confirming that a moderate expansion order is sufficient to capture the bulk of the cosmological information. Isosceles configurations converge more slowly and exhibit a stronger scale dependence, reinforcing the need to treat them separately. These conclusions are supported by the analysis of \citet{EP-Guidi}, who demonstrated that unbiased cosmological constraints can be obtained with $\lmax = 10$ once the problematic configurations are identified and excluded.

For all these reasons, we decided to adopt the SHD as the reference method for the 3PCF estimate and to keep the DC option to perform validation tests and to handle specific, potentially problematic, triangle configurations.

\subsection{Null test}

In this test, we use the SHD method to estimate the 3PCF in the Gaussian mock catalogues described in Sect.~\ref{sec:data}, which are expected to exhibit no 3PCF signal. This test aims to verify that our pipeline can robustly recover a null signal, introducing systematic errors smaller than \SI{10}{\percent} of the statistical error. For this test, we consider Gaussian fields generated on \num{600} grids and we focus on one triangle configuration only, characterised by a single configuration ($r_{12}=40 \, \hMpc$, $r_{13}=80\,\hMpc$) with $\lmax=20$. The number density of the corresponding random catalogue is set to \num{50} times that of the object catalogue, according to the reference value. Each mock has the expected volume and density of the first redshift bin of the EDR1.
The rationale for using \num{600} mocks is to beat the statistical error that, for the mean of the 3PCF measurements, is approximately \num{25} times smaller than that of a single measurement, allowing us to appreciate systematic deviations from the expected null signal as small as \SI{3}{\percent} of the expected statistical error.
The results of the test are shown in Fig.~\ref{fig:gaussian}, where we display the mean signal from the \num{600} mocks (solid black line) compared to the expected statistical error, estimated from the RMS scatter of the \num{600} measured 3PCF (light grey band).
For reference, we also show the \SI{10}{\percent} amplitude of this error (dark grey band).

This test demonstrates the absence of significant systematic effects in estimating the 3PCF. The average signal is consistent with its associated error and thus is well within the scientific requirements. We verified that these results are not specific to the triangle configuration considered but are valid for all the various triangle configurations we explored.

We limited our validation tests to a range of scales in which deviations from Gaussianity are expected to be mild. Performing the null Gaussian test on scales smaller than $20 \, \hMpc$ would be technically challenging, since Poisson noise in the catalogue generation procedure would induce significant deviations from Gaussianity on these scales.

\begin{figure*}[t]
    \centering
    \includegraphics[width=\linewidth]{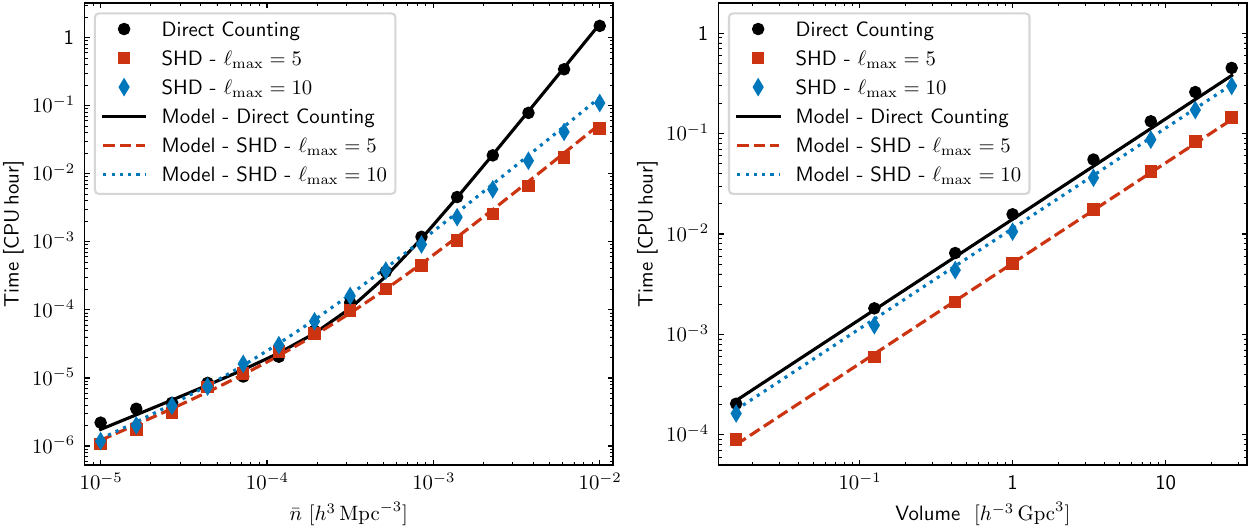}
    \caption{\emph{Left:} Computational cost in CPU--hours as a function of the sample number density for a single triangle configuration ($\rab = 20\,\protect\hMpc$, $\rac = 40\,\protect\hMpc$, $\Delta r = 5\,\protect\hMpc$). Symbols show measured run times; lines show the analytical models from Eqs.~\eqref{eq:t_dir_sin} and \eqref{eq:t_shd_sin}. The DC method (black) scales as $\bar{n}^3$, while the SHD method scales as $\bar{n}^2$, with a weak dependence on $\lmax$ (red: $\lmax = 5$; blue: $\lmax = 10$). \emph{Right:} Same comparison as a function of volume at fixed density. Both methods show linear scaling with volume.}
    \label{fig:triplet_times}
\end{figure*}

\subsection{\label{sec:split}Random splitting}

We introduced the random splitting method \citep{Keihanen2019} in Sect.~\ref{sec:split_description} as a technique to significantly reduce the computational cost of calculating the 3PCF. This technique consists of splitting the random sample into $N_{\rm split}$ independent samples and estimating the 3PCF as described in Eq.~\eqref{eq:zeta_split}. This procedure significantly speeds up computation, since in the process, a significant fraction of the triplets in the parent random catalogue are not included in the counts. In fact, counting triplets in the random sample dominates the computational cost of the procedure. The downside of this procedure is the amplification of shot noise, which is contributed by both the data and the random samples.

To quantify the impact of the split and assess its dependence on the split factor $N_{\rm split}$, defined as the number of subcatalogues into which the parent one was split, we use the Flagship mock catalogue described in Sect.~\ref{sec:data}.
We consider different scenarios characterised by different split factors, defined as the number of independent subsamples into which the parent catalogue was split.
In Fig.~\ref{fig:split-DR1} we show the difference between each 3PCF estimated with each of the split factors considered, indicated on the label, and the reference factor, as a function of the triangle index in units of the expected random error.  The triangle index identifies all triangles with sides in the range from \SIrange[range-phrase={~to~}]{20}{150}{\hMpc}, excluding the isosceles.

From the plot, it is evident that even in the most extreme case ($N_{\rm split}=50$), the error introduced by the split is consistently below the requirements ($\sigma_{\rm syst} <
\SI{10}{\percent} \,\sigma_{\rm stat}$), and increasing with the density of the random samples decreasing (proportional to the number of subsamples). This test demonstrates the robustness of this method, allowing us to opt for large split values to maximise performance without losing accuracy. \citet{Slepian2015} and \citet{Keihanen2019} showed that for the 3PCF, the optimal density of the random sample is $1.5 N_{\rm g} < N_{\rm r} < 2 N_{\rm g}$. For the case of \Euclid, we conservatively choose $N_{\rm split} =$ \num{25} as the standard. As shown below, the time gain saturates for this split value, making more aggressive solutions unnecessary.

\subsection{\label{sec:edgecorr} Edge correction}

The Szapudi--Szalay estimator introduced in Eq.~\eqref{eq:SS98_expanded} includes the correction for edge effects and proved effective for this type of analysis with the precision required for \Euclid. However, to use this estimator with the SHD method, we should first reconstruct the triplet counts from harmonic space to triangle space (Direct Sum, Eq.~\ref{eq:zeta_ds}). Alternatively, we can use the estimator proposed by \citet{Slepian2015} which provides the 3PCF directly in harmonic space, and then get the reconstructed 3PCF ($\zeta_{\rm SE}$, Eq.~\ref{eq:zeta_se}).  To appreciate the impact of the edge effects, we compared the results obtained with the two methods. To quantify the difference between the two methods, we measured the 3PCF for the Flagship sample described in Sect.~\ref{sec:data}, on scales $20 < r/\hMpc < 150$, in bins of \SI{10}{\hMpc} and up to $\lmax=10$. In Fig.~\ref{fig:edgecorr}, we show the difference between the two estimates of $\zeta$ in the expected statistical error units. Each point represents the 3PCF value for a particular triangle, with colours indicating the ratio $r_{13}/r_{12}$, from isosceles configurations ($r_{13}/r_{12}=1$) to more elongated configurations. The size of the triangle sides increases from left to right.

The differences between the two methods amount to less than \SI{3}{\percent} of the expected statistical error and will not affect the error budget. For reference, the grey band in Fig.~\ref{fig:edgecorr} marks $\pm\SI{1}{\percent}$ of the statistical error, an order of magnitude below the \SI{10}{\percent} requirement.  Although this result may depend on the footprint of the survey, the smallness of these errors makes us confident that edge effects in the real survey will not significantly impact the total error budget.  However, it is interesting that, although small, the discrepancy between the two methods increases with the size of the triangle (which correlates with the triangle index) when this becomes comparable with the size of the survey. As expected, the effect increases with scale since the random corrections are more important at large scales, comparable with the size of the volume sampled. There is also a slight trend with the triangle shape, with more pronounced differences for isosceles or nearly isosceles configurations. The results were obtained for this measurement choice, with $\lmax=10$. Higher values of $\lmax$ allow us to achieve better agreement. However, the choices considered in this analysis seem sufficient to obtain a robust measurement.

The reason for the insensitivity to edge effects is due to the fact that in harmonic space, the signal related to the triplets of the random catalogue converges very quickly and is sizable only for the first few multipoles \citep{Slepian2015}. This result depends on the survey geometry and should be tested for various usage scenarios.

\section{\label{sec:performances} Computational performance}
The goal of the \Euclid survey is to measure the correlation properties of tens of millions of objects. This task, using the estimators adopted for two- and three-point correlation analyses, relies on random samples of synthetic objects that are up to \num{50} times larger than the galaxy sample itself. Even with highly optimised estimators like those presented here, this task remains computationally challenging. Therefore, accurate estimation of computational requirements is essential for planning a sustainable analysis within the constraints of the available computational facilities.

This is especially true because these analyses will need to be repeated multiple times to process different data sets, which may be either real (e.g., galaxy catalogues selected based on various criteria) or simulated (as numerous independent 3PCF measurements are required to construct a reliable covariance matrix).

In this section, we focus on quantifying the computational performance of the \Euclid library to calculate the 3PCF.
First, we introduce models for computational time in triplet-counting algorithms. These models encapsulate the basic behaviours of different triplet counting algorithms described in previous sections with respect to fundamental quantities such as the number density $\bar{n}$, the volume $V$, the triangular configurations, and the value of $\lmax$.
After calibrating and testing these models against actual code runs, we will then evaluate (i) the impact of the parallelisation procedure on computational
time and its efficiency, and (ii) the performance improvement resulting from the introduction of
the random sample split method, described in Sect.~\ref{sec:random_split}.
We will then quantify the impact of computational resources scaled to the \Euclid survey. In Appendix\,\ref{sec:appC}, we will compare our results with public 3PCF codes available in the literature.
All computational tests presented in this section were performed on a dual-socket AMD EPYC 7413 system, providing 48 physical cores (96 threads) at a base frequency of \SI{2.65}{\giga\hertz} and \SI{256}{\giga\byte} of RAM.

\subsection{\label{sec:cost} Computational cost}

The triplet counting procedure undoubtedly dominates the computational cost of the 3PCF. From an analytical point of view, it is easy to write the relation between the total time and the type of measurement. However, it is necessary to differentiate the four scenarios already described above.

The formulas presented in the following pertain only to the triplet counting procedure and do not consider the entire process necessary for estimating the 3PCF.

\subsubsection{Direct Counting}
The brute force version of the algorithm should formally scale as $\mathcal{O}(N^3)$. The linked list partitioning reduces the number of operations needed to complete the computation, which now becomes the number of particles times the search volume squared ($t \propto \bar{n} ^3 V r_{\rm max}^6$).

For the single configuration case, we have
\begin{equation}
    \label{eq:t_dir_sin}
    t \, = \, 2 \times 10^{-10} \, V_{12} \,  V_{13} \, \left( \frac{\bar{n}}{10^{-3}} \right)^3 \, \left(\frac{V}{V_{\rm DR3}}\right) \, \, [\text{CPU\,hour}]\;,
\end{equation}
where $V_{12}, \, V_{13}$ are the volumes of the spherical shells in which the triplet counting occurs. In Eqs.~\eqref{eq:t_dir_sin}--\eqref{eq:t_shd_all}, $\bar{n}$ is in $h^3\,\mathrm{Mpc}^{-3}$, volumes ($V$, $V_{12}$, $V_{13}$) are in $h^{-3}\,\mathrm{Mpc}^3$, and $r_{\rm max}$ is in $h^{-1}\,\mathrm{Mpc}$; the normalising constants ($10^{-3}$, $150$, \textit{etc.}) carry the same units as the quantities they scale, so that each bracket is dimensionless.
For the all configuration case, we have
\begin{equation}
    \label{eq:t_dir_all}
    t \, = \, 4 \times 10^{4} \, \left( \frac{\bar{n}}{10^{-3}} \right)^3 \, \left(\frac{V}{V_{\rm DR3}}\right) \, \left( \frac{r_{\rm max}}{150}\right)^6 \, \, \, \, [\text{CPU\,hour}]\;.
\end{equation}
Here, $r_{\rm max}$ is the maximum separation considered to count the triplets.

\subsubsection{Spherical Harmonics Decomposition}
Similarly to DC, the linked-list helps reduce its formal scaling $t \propto \mathcal{O}(N^2)$ to the number of particles times the search volume, $t \propto \bar{n}^2 V r_{\rm max}^3$.
As explained in Sect.\,\ref{sec:SHD}, the SHD method depends on the choice of the maximum multipole index $\lmax$.
When considering the computational budget for the SHD triplet counting, we need to introduce an explicit dependence on this parameter.

For the single-configuration case, we have
\begin{equation}
        \label{eq:t_shd_sin}
        t \, = \, 3.0 \times 10^{-6} \, \left(V_{12} + V_{13} \right) \,
        \left( \frac{\bar{n}}{10^{-3}} \right)^2 \, \left(\frac{V}{V_{\rm DR3}}\right) \, \left(\frac{\lmax}{10}\right)^{1.5} \, \, [\text{CPU\,hour}] \;,
\end{equation}
where we have made explicit the dependence on $\lmax$, expressing it as a power law.
For the all configurations case, we have
\begin{equation}
    \label{eq:t_shd_all}
    t \, = \, 50 \left( \, \frac{\bar{n}}{10^{-3}} \right)^2 \, \left(\frac{V}{V_{\rm DR3} }\right) \, \left( \frac{r_{\rm max}}{150}\right)^3 \, \left(\frac{\lmax}{10}\right)^{1.5} \, \, [\text{CPU\,hour}] \;.
\end{equation}

These simple formulas show asymptotic trends for high-density regimes, valid for the densities expected from the \Euclid survey. We notice that the difference between the DC and SHD methods is huge, which confirms the second method as our preferred choice.
The proportionality constants were determined by directly comparing these models to actual measurements of 3PCF in different configurations. As an example, in the left panel of Fig.~\ref{fig:triplet_times} we report the computational times in CPU hours for the case of 3PCF in single configuration with $r_{12} = 20 \, \hMpc$, $r_{13} = 40 \, \hMpc$, $\Delta r = 5 \, \hMpc$, varying the density of the sample generated in a box with a side length of $500 \, \hMpc$, while in the right panel we compute the 3PCF in the same cases but for different volumes at fixed density. We compare the models in Eqs.~\eqref{eq:t_dir_sin} and \eqref{eq:t_shd_sin}, shown with lines, and the actual performance of our algorithm, shown with symbols. In particular, in black, we report the scaling for the case of direct triplet counting, while in red and blue, we show the times related to the SHD technique for $\lmax=5, 10$, respectively, in red and blue. The predictions accurately describe the trends as a function of density. In the shown curves, we also included a term to reproduce the trends in low-density regimes, where, due to the scarcity of sources, other steps of the algorithms dominate the computational cost, changing the slope of the relationship. The main contribution is the creation of the linked list, which scales as $\mathcal{O}(N)$ and becomes predominant at these very low densities where the triplet counting with this method becomes highly efficient. We do not explicitly include this contribution in the above formulas, as it is only significant in unrealistic regimes, significantly sparser than the number densities expected for the \Euclid survey.
We are therefore confident that we can use these relationships to predict the computational times in the case of our interest, as we will do in the following sections.

\subsection{\label{sec:parallelisation} Parallelisation}

\begin{figure}[t]
    \centering
    \includegraphics[width=\linewidth]{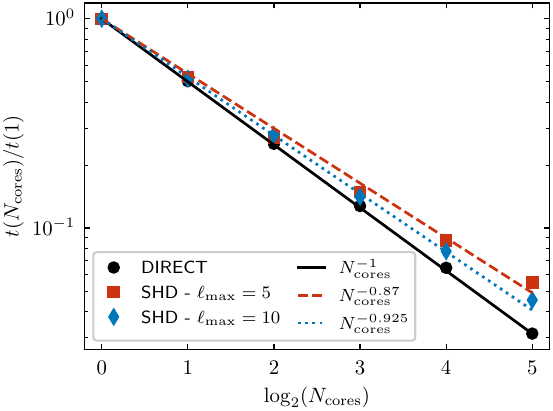}
    \caption{Trend of computational cost as a function of the number of threads, compared to the single-core case for the two triplet counting methods. The black dots represent the computational time measurements for the DC case, while the SHD case is represented with red squares ($\lmax=5$) and blue diamonds ($\lmax=10$). The measurements were obtained by running the algorithm in single configuration mode with $r_{12}=20\,\protect\hMpc$, $r_{13}=40\,\protect\hMpc$, $\Delta r=5\,\protect\hMpc$, and at a fixed density $\bar{n} = 2 \times 10^{-3} \, h^3\,\mathrm{Mpc}^{-3}$.}
    \label{fig:time_vs_threads}
\end{figure}

An effective parallelisation is of paramount importance to efficiently distribute the computational burden. As illustrated in Sect.~\ref{sec:triplets}, triplet counting is a procedure performed iteratively using the position of each galaxy in the catalogue as the primary vertex. This procedure can be effectively parallelised by distributing computations related to the individual primary galaxies across multiple threads.
From a technical point of view, parallelisation is implemented using \texttt{OpenMP}, as required by the working environment in which this code is developed. This implies that parallelisation can occur across multiple threads of the same node, but not across different nodes. In some cases, such as splitting the random sample, this limitation can be circumvented, further reducing the computational cost, provided that computational resources are available.
In this section, we test the efficiency of parallelisation by comparing the scaling of the code with the number of threads. Figure~\ref{fig:time_vs_threads} shows the scaling of the triplet counting algorithm with the number of threads. In particular, we report, with symbols, the computational time of different runs normalised to the time required to run on a single thread. We used the same configuration as the one reported in the previous section for these estimates. This test demonstrates that the scaling in the case of DC counting (black solid line, black dots) is ideal and follows $N_{\rm threads}^{-1}$. The SHD case (dashed line and red squares, $\lmax=5$, dotted line and blue diamonds, $\lmax=10$) is slightly less efficient, with a scaling of around $N_{\rm threads}^{-0.9}$, and shows greater efficiency as a function of the value of $\lmax$. However, despite being more effectively parallelised, the DC method remains much more computationally demanding than the SHD one.
As a practical example, for the test configuration used in this section, the DC method on 32 threads is approximately 32 times faster than on a single core, while the SHD method achieves a factor of $\sim 24$ speed-up. Despite being more efficiently parallelised, the DC method remains orders of magnitude slower than SHD for all-configuration runs at \Euclid-like densities (see Sect.~\ref{sec:Euclid_times}).

\subsection{\label{sec:random_split} Computational gain of the random split}

\begin{figure}[t]
    \centering
    \includegraphics[width=\linewidth]{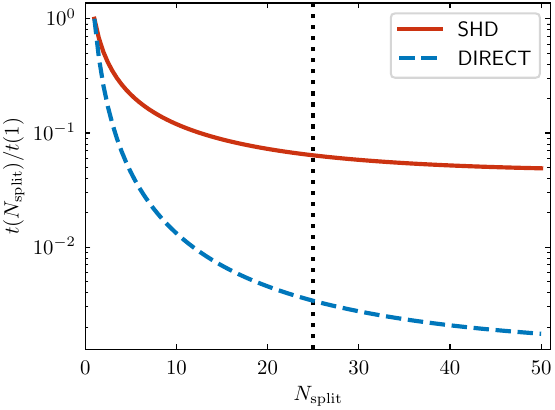}
    \caption{Computational time as a function of the split factor $N_{\rm split}$, normalised to the no-split case, for the DC (blue dashed) and SHD (red solid)  methods. The vertical dashed line marks the reference choice $N_{\rm split} = 25$. The split yields a factor of ${\sim}\,10$ gain for SHD and ${\sim}\,500$ for DC.}
    \label{fig:full_zeta_split}
\end{figure}

\begin{figure*}[t]
    \includegraphics[width=\textwidth]{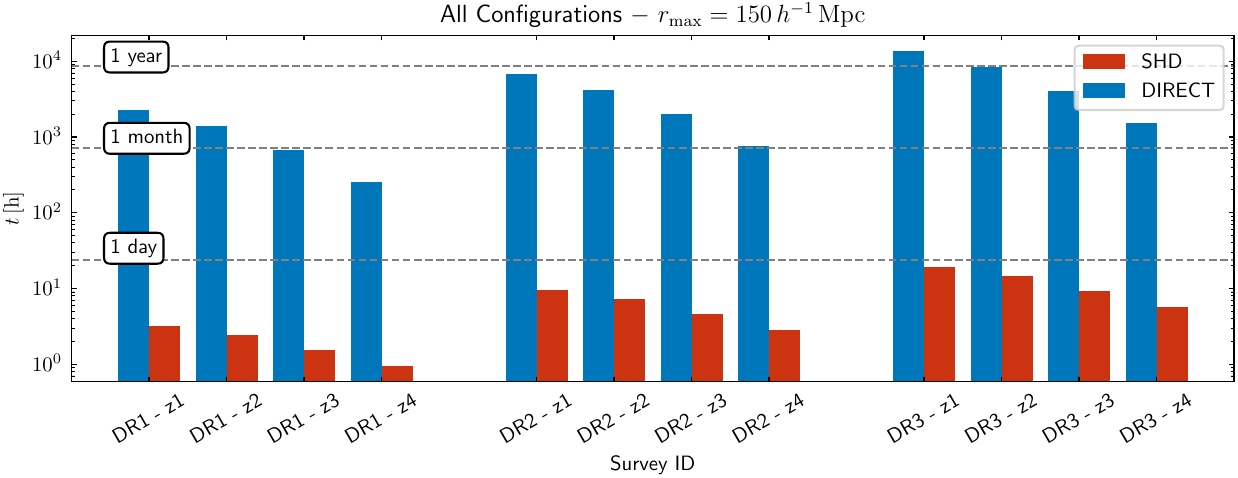}
    \caption{Predicted computational times for measuring the 3PCF across all configurations up to $r_{\rm max} = 150\,\protect\hMpc$ for the three \Euclid data releases and four redshift bins. Predictions assume a random sample \num{50} times denser than the galaxy catalogue, a split factor of $25$, $32$ threads, and a clock frequency of $3\,\mathrm{GHz}$, using the survey parameters from Table~4 of \citet{Blanchard-EP7}. The DC method (blue) requires months to years, while the SHD method (red) completes within a day even for the largest data releases, confirming it as the only viable strategy for the \Euclid spectroscopic survey.}
    \label{fig:computational_budget}
\end{figure*}

So far, we focused on the triplet counting part of the algorithm, whose computational impact was mitigated by parallelising the code and by adopting data partitioning schemes like the linked-list. To reduce the computational cost of the remaining part of the algorithm, in which the counts are combined to estimate the 3PCF, we adopted the split method introduced in Sect.~\ref{sec:split_description}. We quantify the computational gain of its adoption and its dependence on the split factor with the following relation
\begin{align}
    \label{eq:zeta_time}
    t_{\zeta}  =  \frac{N_{\rm split}}{N_{\rm threads}} & \left\{ t_{\rm NNN}\left(\bar{n}_{\rm g} + \bar{n}_{\rm g} \, \frac{N_{\rm R}}{N_{\rm split}}; r_{\rm max}, \,\lmax\right) \right. \nonumber \\
    & + \left. t_{\rm RRR}\left(\bar{n}_{\rm g} \, \frac{N_{\rm R}}{N_{\rm split}}; r_{\rm max}, \, \lmax\right) \right\} \;.
\end{align}
We assume that the random sample is always \num{50} times denser than the galaxy catalogue and similar to Sect.~\ref{sec:split}, we consider it either as a whole ($N_{\rm split}=1, N_R=50$) or divided up to $N_{\rm split}=50$.
The formulas for the computational time of the triplet counts, $t_{\rm NNN}$ and $t_{\rm RRR}$ are those reported in Eqs.\,\eqref{eq:t_dir_sin}--\eqref{eq:t_shd_all} depending on the case under consideration, to be filled with all other parameters that control the total triplet time, such as the maximum separation of the shells, $r_{\rm max}$, or the value of $\lmax$ in the case of the SHD method.
In Fig.~\ref{fig:full_zeta_split}, we plot the trends of the computational time, relative to the no-split case, as a function of the split parameter for the DC and SHD methods, shown respectively with a dashed blue line and a solid red line. These trends are obtained starting from Eq.\,\eqref{eq:t_dir_all} and Eq.\,\eqref{eq:t_shd_all}, changing the values of $\bar{n}$ accordingly, and rescaling for the number of splits. The predictions show how the split can lead to an efficiency gain of a factor of 10 (500) for the SHD (DC) case. The difference in efficiency between the two methods is due to the difference in behaviour with density. Although the DC method benefits more prominently from the split, it should be emphasised that the absolute computational time is still $10^3$ times higher compared to the SHD case.
By equating Eqs.~\eqref{eq:t_dir_all} and \eqref{eq:t_shd_all}, the crossover density at which the DC and SHD methods have equal computational cost scales as $\bar{n}_{\rm cross} \propto (\lmax/10)^{1.5} \, (r_{\rm max}/150)^{-3}$. For $r_{\rm max} = 150\,\hMpc$ and $\lmax = 10$, this gives $\bar{n}_{\rm cross} \sim 10^{-6}\,h^3\,\mathrm{Mpc}^{-3}$, well below the expected \Euclid density. At $\bar{n} = 6 \times 10^{-4}\,h^3\,\mathrm{Mpc}^{-3}$, the SHD method is approximately 500 times faster than DC for the all-configuration case.

\subsection{\label{sec:Euclid_times} Computation forecast for \Euclid spectroscopic survey}
Finally, we provide an estimate of the expected computing time required to measure the 3PCF using both the DC and SHD methods for the various galaxy catalogues that will be extracted from the spectroscopic survey in the three public data releases.
To obtain these estimates, we use Eq.~\eqref{eq:zeta_split}, which, as discussed, depends on several parameters set as follows. The parameters that define the properties of the samples, such as area, redshift interval, volume, and number density, are set to the values listed in Table 4 of \citet{Blanchard-EP7}. For the parameters defining the three-point clustering analysis, we consider triangles with a maximum size of $r_{\rm max}=150 \, \hMpc$, a random catalogue $50$ times denser than the galaxy catalogue, and a split factor $N_{\rm split}=25$. For the SHD method, we compute up to $\lmax=10$ multipoles. Finally, we assume the computations are performed on 32 threads.
The predictions are shown in Fig.~\ref{fig:computational_budget} and demonstrate how the SHD method, shown in red, is extremely competitive, allowing the measurement in times on the order of a day. However, the direct counting method, shown in blue, is highly costly and is not feasible for this type of analysis.
These same results are mitigated in the case of single configuration and allow us to consider the DC method for isosceles configurations where we know the SHD method struggles to converge. The scientific relevance of these cases will need to be evaluated appropriately and is beyond the scope of this paper.
Finally, we note that these conclusions are relative to the entire spectroscopic galaxy survey. The results could be less drastic when used with sparser samples, such as galaxy cluster catalogues, or over smaller volumes, as in the case of the Euclid Deep Survey (EDS).

\section{\label{sec:conclusions} Conclusions}
This paper presents the numerical code that was developed and implemented in the SGS data analysis pipeline to measure the galaxy 3PCF of the \Euclid spectroscopic galaxy survey. Like all the other codes that will be used to measure clustering statistics, the 3PCF code has to adhere to stringent requirements on both precision and accuracy to guarantee that the evaluation process has a negligible impact on the error budget of the clustering analysis.
To obey this constraint, the 3PCF code implements state-of-the-art methods for measuring the isotropic 3PCF across the range of scales selected by the user in the triangle space. The code incorporates the Szapudi--Szalay estimator and is based on triplet counts. These triplets can be estimated by direct counting or in harmonic space.
The default method that we adopt for the \Euclid data analysis is the one based on SHD. The main reason for this choice is computational cost, which must be small enough to estimate the 3PCF of the final \Euclid data set. The reduced computational cost comes at the expense of some loss in accuracy; for non-isosceles configurations with $\lmax = 10$, the difference between the SHD and DC estimates is well within the expected statistical error (see Fig.~\ref{fig:iso_noniso}), meeting the scientific requirement of a systematic bias below \SI{10}{\percent} of the expected statistical uncertainty. Isosceles configurations, for which convergence is significantly slower, will need to be handled differently.
To assess the adequacy of our 3PCF code and to gauge its performance, we designed and performed several scientific and numerical tests.

First of all, to evaluate the precision and the accuracy of the code we performed a null test in which we estimated the 3PCF of a set of objects extracted from a Gaussian realisation, in which, by definition, the intrinsic 3PCF is zero. Since the target accuracy is as high as \SI{10}{\percent} of the expected statistical error, we repeated the measurement on \num{600} simulated catalogues and then averaged to keep the error in the mean small enough. The number density of the objects in these catalogues was set equal to $6\times10^{-4} \, h^{3} \, \mathrm{Mpc}^{-3} $ to match the expected density of the spectroscopic survey of \Euclid at $z=1$.
The test verifies whether the null hypothesis holds when the 3PCF is measured in many of these mocks. To achieve stable measurements, the signal was averaged over \num{600} mocks. We built these mocks to have the same volume and number density as EDR1. The results confirm that the systematic error is within acceptable limits, namely \SI{10}{\percent} of the statistical error.
In the second test, we evaluated the sensitivity of the accuracy of the SHD method to the choice of the maximum multipole $\lmax$, using the corresponding DC result as a reference. The goal was to find the best trade-off between accuracy and computational cost, as they both increase with $\lmax$. We verified that the 3PCF obtained by summing the multipoles $\zeta_{\ell}$ converges to the DC result for a moderate value $\lmax=10$ for non-isosceles triangle configurations. Convergence is much slower for isosceles configurations ($r_{12}=r_{13}$). Moreover, the value at which it is reached depends on the specific configuration, making it difficult to choose a value for $\lmax$ that is adequate for all cases and, at the same time, small enough to be computationally manageable.
This difficulty of the SHD method to handle the isosceles configurations does not necessarily have a negative impact on cosmological inference, which can be based on the multipoles $\zeta_{\ell}$ rather than on the full 3PCF.
The remarkable computational performance of our code also relies on the use of the split method to estimate the 3PCF from the counts. This approach allows us to speed up the computation by a factor of 10 (\num{100}) for the SHD (DC) without significantly amplifying random errors or introducing systematic ones, as we verified in our third validation test.

The fourth test was in fact a challenge between various publicly available codes to estimate the galaxy 3PCF, confirming the robustness and competitiveness of the \Euclid 3PCF processing element. These comparisons validate the excellent agreement, to machine precision, between the \Euclid 3PCF code and other existing tools, while also highlighting its competitive computational performance.
The code presented here can be further developed to analyse future \Euclid catalogues that, due to their increasing size, will allow us to extend the 3PCF analysis beyond the purely isotropic one. The first natural direction is the extension to anisotropic 3PCF \citep{Slepian2018}, which can be achieved with minimal extension. A further development direction is the code extension for the computation of the $N$-point correlation function, obtained by generalising the SHD formalism.
This will allow \Euclid to have a powerful toolkit to explore clustering, deeply connected to the data reduction pipeline, to improve the extraction of suitable information for cosmological analysis.

\begin{acknowledgements}
\AckEC
\AckCosmoHub
\end{acknowledgements}

\bibliography{Euclid, DR1, refs}

\begin{appendix}

\section{\label{sec:appA} Binned Legendre multipoles}
The direct method is based on counting triplets in bins with a predetermined size. For a fair comparison with the SHD method, it is necessary to re-sum the triplets in harmonic space, taking into account both the bin size and the chosen type of parameterisation (see Eq.~\ref{eq:zeta_ell_binned_def}).
In case of {\small COSTHETA}
parameterisation, the expression for $\bar{\legG}$
is trivial, and reads
\begin{equation}
    \label{eq:leg_ave_mu}
    \bar{\legG} \, = \, \frac{1}{\Delta \mu} \int_{\mu_{\rm min}}^{\mu_{\rm max}} \diff \mu' \legG(\mu') \, = \, \left[ \frac{\mathcal{L}_{\ell+1}(\mu) - \mathcal{L}_{\ell-1}(\mu)}{2\ell+1}\right]_{\mu_{\rm min}}^{\mu_{\rm max}}\;,
\end{equation}
where $\mu_{\rm min}, \mu_{\rm max}$ are the bin edges in $\mu$.
A similar expression can be used for {\small THETA} parameterisation.

As already mentioned, the {\small SIDE} case is more complex, requiring a multidimensional integration over $r_{12}, r_{13}$ bins.
It is possible to demonstrate that the average Legendre multipoles can be written as
\begin{multline}
    \bar{\legG}\left( r_{12}, r_{13}, r_{23} \right) = \\
    \frac{\displaystyle\int_{k_{\rm min}}^{k_{\rm max}}  \diff k \, k^2 \, \mathcal{I}_{\ell}(k; \Delta r_{12}) \mathcal{I}_{\ell}(k; \Delta r_{13}) \mathcal{I}_0(k ; \Delta r_{23})}{\displaystyle\int_{k_{\rm min}}^{k_{\rm max}}  \diff k \, k^2 \, \mathcal{I}_0(k; \Delta r_{12}) \mathcal{I}_0(k; \Delta r_{13}) \mathcal{I}_0(k; \Delta r_{23})}\;,
\end{multline}
where the integrals are evaluated over the range $k_{\rm min} = 10^{-4}\,h\,\mathrm{Mpc}^{-1}$ to $k_{\rm max} = 10^{2}\,h\,\mathrm{Mpc}^{-1}$, $\Delta r_{12}, \Delta r_{13}, \Delta r_{23}$ are the limits for the three sides respectively, and $\mathcal{I}_{\ell}$ is the spherical Bessel function of order $\ell$ averaged over the spherical shell of volume $V$:
\begin{equation}
    \mathcal{I}_{\ell} (k; \Delta r) = \frac{4 \pi \int_{r_{\rm min}}^{r_{\rm max}} \diff r \, r^2 \, j_{\ell} (k r)}{V}\;.
\end{equation}
However, it is important to note that the expression in Eq.~\eqref{eq:zeta_ell_binned_def} is useful for comparison with the direct method and efficient compression. When using the 3PCF in harmonic space, the type of re-summation can be easily absorbed into the modelling, which can be more easily dealt with in harmonic space. The advantage of resummation lies in the compression of $\zeta_{\ell}$ when expanded over a large number of Legendre polynomials, helping to compute covariance, and in the ability to make a more refined selection of scales in likelihood analysis.

\section{\label{sec:appB} Theoretical covariance}
We use the expression presented in section~6 of \citet{Slepian2015} to calculate the analytical covariance matrix of the 3PCF. There the authors describe the covariance for the Legendre coefficients of the 3PCF
\begin{equation}
\begin{aligned}
    \label{eq:AnCov}
    \text{Cov}_{\ell \ell'}(r_1, r_2; r_1', r_2') &= \frac{4\pi}{V} (2\ell+1)(2\ell'+1)(-1)^{\ell+\ell'} \\
    &\quad \times \int r^2 \, \diff r \sum_{\ell_2} (2\ell_2+1)
    \begin{pmatrix}
    \ell & \ell' & \ell_2 \\
    0 & 0 & 0
    \end{pmatrix}^2 \\
    &\quad \times \Bigg\{ (-1)^{\ell_2} \xi_0(r)
    \Bigg[ f_{\ell_2\ell\ell'}(r; r_1, r_1') f_{\ell_2\ell\ell'}(r; r_2, r_2')\\
    &\quad + f_{\ell_2\ell\ell'}(r; r_2, r_1') f_{\ell_2\ell\ell'}(r; r_1, r_2') \Bigg] \\
    &\quad+ (-1)^{(\ell+\ell'+\ell_2)/2} \\
    &\quad\times \Bigg[ f_{\ell\ell'}(r; r_1) f_{\ell\ell'}(r; r_1') f_{\ell_2 \ell \ell'}(r; r_2, r_2') \\
    &\quad+ f_{\ell\ell'}(r; r_2) f_{\ell\ell'}(r; r_2') f_{\ell_2\ell\ell'}(r; r_1, r_1') \\
    &\quad + f_{\ell\ell'}(r; r_1) f_{\ell\ell'}(r; r_2') f_{\ell_2\ell\ell'}(r; r_2, r_1') \\
    &\quad + f_{\ell\ell'}(r; r_2) f_{\ell\ell'}(r; r_1') f_{\ell_2 \ell \ell'}(r; r_1, r_2') \Bigg] \Bigg\}\;,
\end{aligned}
\end{equation}
where the integral is evaluated from $r_{\rm min} = 0 \, \protect \hMpc$ to $r_{\rm max}=1000 \, \protect \hMpc$.
Here $\xi_0$ is the Hankel transform of the power spectrum $P(k)$, and the terms $f_{\ell\ell'}$ and $f_{\ell_2\ell\ell'}$ are its one-dimensional integrals
\begin{align}
    \label{eq:f_ells}
    f_{\ell\ell'}(r; r_1) & = \int_{k_{\rm min}}^{k_{\rm max}} \frac{k^2 \, \diff k}{2\pi^2} P(k) j_{\ell}(kr_1) j_{\ell}(kr)\;, \\
    f_{\ell_2\ell\ell'}(r; r_1, r_2) & = \int_{k_{\rm min}}^{k_{\rm max}} \frac{k^2 \, \diff k}{2\pi^2} P(k) j_{\ell}(kr_1) j_{\ell'}(kr_2) j_{\ell_2}(kr)\;,
\end{align}
with the same $k_{\rm min}$ and $k_{\rm max}$ as in Appendix~\ref{sec:appA}.

In our case, we chose to represent the 3PCF in the space of triangles. To obtain the covariance on this basis, we apply the transformation

\begin{equation}
    \label{eq:cov_resum}
    {\rm Cov}(r_{1}, r_{2}, \mu, r_{1}^{\prime}, r_{2}^{\prime}, \mu^{\prime}) \, = \, \sum_{\ell, \ell^{\prime}} {\rm Cov}_{\ell \ell^{\prime}}(r_{1}, r_{2}, r_{1}^{\prime}, r_{2}^{\prime}) \mathcal{L}_{\ell}(\mu) \mathcal{L}_{\ell^{\prime}}(\mu^{\prime}) \;.
\end{equation}

\section{\label{sec:appC} Comparison with external codes}
\begin{table*}
    \caption{Comparison of various cosmological libraries from the literature with the official \Euclid code. The comparison is limited to functionalities related to the estimate of the 3PCF and related aspects used in this paper. Green symbols highlight codes that perform slightly better than \Euclid while red ones indicate those performing significantly worse.}
    \renewcommand{\arraystretch}{1.7}

    \centering
    \begin{tabular}{|c|c|c|c|c|c|c|}
    \hline
      \multirow{2}{*}{Library}  & \multicolumn{2}{c|}{Direct Counts} & \multicolumn{2}{c|}{Harmonic space} & \multirow{2}{*}{Extra} & \multirow{2}{*}{Reference} \\
      \cline{2-5}
               & Single & All & Single & All& & \\
    \hline
       \Euclid  & \ding{51} & \ding{51} & \ding{51} & \ding{51} & \Euclid archives I/O & This work \\
    \hline
       \texttt{CosmoBolognaLib} & \;\;\;\;\; \ding{51}(\textcolor{red}{$\times 5$}) & \ding{55} & \;\;\;\;\;\ding{51}(\textcolor{red}{$\times3$}) & \;\;\;\;\;\ding{51}(\textcolor{red}{$\times 2$}) & Jackknife/Bootstrap & \citet{Marulli2016} \\
    \hline
       \texttt{ENCORE} & \ding{55} & \ding{55} & \ding{55} & \;\;\;\;\;\;\; \ding{51} (\textcolor{OliveGreen}{$\times 0.9$}) & PBC, AVX, GPU, NPCF & \citet{Philcox2021}\\
    \hline
       \texttt{MeasCORR} & \ding{55} & \ding{55} & \ding{55} & \;\;\;\;\;\ding{51} (\textcolor{red}{$\times2$}) & PBC, Anisotropic 3PCF & \citet{Farina2026} \\
    \hline
    \end{tabular}\vspace{0.25cm}

    \label{tab:code_comparison}
\end{table*}

Our code is seamlessly integrated into the \Euclid data analysis pipeline and is highly specialised for its tasks. Given the recent interest in higher-order statistics, mainly the 3PCF, the literature presents several competitive solutions for extracting this measure, along with a suite of complementary tools.

In this Appendix, we compare the computational performance of the \Euclid code with three publicly available software available in the literature. The selected software are \texttt{CosmoBolognaLib} \citep{Marulli2016}, \texttt{ENCORE} \citep{Philcox2021}, and \texttt{MeasCorr} \citep{Farina2026}. These tools were chosen because they feature characteristics that are different and complementary to those of our code. In the following, we briefly describe the relevant features of these codes. We encourage readers to refer to the specific papers for detailed implementation descriptions.

\begin{description}
\item[\texttt{\bf CosmoBolognaLib}:]
The \texttt{CosmoBolognaLib}\footnote{\url{https://gitlab.com/federicomarulli/CosmoBolognaLib}} \citep{Marulli2016} is a free software library written in {\small C++} and {\small Python}, designed for comprehensive cosmological data analysis, from initial data processing to advanced modelling. It includes tools for measuring and modelling two-point and three-point clustering statistics, galaxy cluster count analysis, void analysis, and more. For the 3PCF, this library implements the DC and SHD method. The pair and triplet counting routines are optimised with the linked list and parallelised using \texttt{OpenMP}. It also estimates the 3PCF covariance matrix from the input catalogue itself using the Jackknife and Bootstrap methods \citep{Norberg2009}.

\item[\texttt{\bf ENCORE}:]
\texttt{ENCORE}\footnote{\url{https://github.com/oliverphilcox/encore}} \citep{Philcox2021} is a public library written in {\small C++} and {\small Python} to calculate isotropic $N$-point correlation functions on all scales up to a maximum separation set by the user, up to $N=6$. The implemented algorithm, which is based on the generalisation of the method presented in \citet{Slepian2015}, is highly efficient with a complexity $O(N^2)$ to which a necessary overhead must be added. The code is parallelised using \texttt{OpenMP} and offers additional optimisation features such as compatibility with \texttt{AVX} vectorisation and the possibility to run GPUs. We do not enable any of these optimisation features while performing our code comparison.

\item[\texttt{\bf MeasCorr}:]
\texttt{MeasCorr}\footnote{\url{https://gitlab.com/veropalumbo.alfonso/meascorr}} is a public library written in {\small C++} and in {\small Python}, focused on measuring the multipoles of two-point and three-point anisotropic correlation functions \citep{Farina2026}. For the 3PCF, the estimator implements harmonic space expansion \citep{Slepian2015} in the isotropic case and tripolar space expansion in the anisotropic case \citep{Sugiyama2019}, with the flexibility to expand to any order. The library is optimised for working with simulated and light-cone data and is easily expandable for measuring higher-order statistics.
\end{description}

In Table \ref{tab:code_comparison}, we summarise the features of interest implemented in the various codes and compare them with the \Euclid code regarding computational performance. The \Euclid code is competitive and performs similarly or better than the current state-of-the-art. Furthermore, when run with identical setups, the code outputs are identical to machine precision, which confirms the perfect agreement of the results obtained for \Euclid.

\end{appendix}

\label{LastPage}
\end{document}